\documentclass[aps,prl,twocolumn,superscriptaddress,floatfix,longbibliography]{revtex4-1}
\usepackage{graphicx}
\usepackage{color}
\usepackage{float}
\usepackage{rotating}

\newcommand{\dI}{$\Delta I/I$}
\newcommand{\dR}{$\Delta R$/$R$}
\newcommand{\dIp}{$\Delta I_p$}

\begin{document}

\title{Using ultrashort optical pulses to couple ferroelectric and ferromagnetic order in an oxide heterostructure}

\author{Y. M. Sheu}
\affiliation{Center for Integrated Nanotechnologies, MS K771, Los Alamos National Laboratory, Los Alamos, New Mexico 87545, USA}
\author{S. A. Trugman}
\affiliation{Center for Integrated Nanotechnologies, MS K771, Los Alamos National Laboratory, Los Alamos, New Mexico 87545, USA}
\author{L. Yan}
\affiliation{Center for Integrated Nanotechnologies, MS K771, Los Alamos National Laboratory, Los Alamos, New Mexico 87545, USA}
\author{Q. X. Jia}
\affiliation{Center for Integrated Nanotechnologies, MS K771, Los Alamos National Laboratory, Los Alamos, New Mexico 87545, USA}
\author{A. J. Taylor}
\affiliation{Center for Integrated Nanotechnologies, MS K771, Los Alamos National Laboratory, Los Alamos, New Mexico 87545, USA}
\author{R. P. Prasankumar}
\affiliation{Center for Integrated Nanotechnologies, MS K771, Los Alamos National Laboratory, Los Alamos, New Mexico 87545, USA}

\date{\today}

\begin{abstract}
A new approach to all-optical detection and control of the coupling between electric and magnetic order on ultrafast timescales is achieved using time-resolved second harmonic generation (SHG) to study a ferroelectric (FE)/ferromagnet (FM) oxide heterostructure. We use femtosecond optical pulses to modify the spin alignment in a Ba$_{0.1}$Sr$_{0.9}$TiO$_{3}$(BSTO)/La$_{0.7}$Ca$_{0.3}$MnO$_{3}$  (LCMO) heterostructure and selectively probe the ferroelectric response using SHG. In this heterostructure, the pump pulses photoexcite non-equilibrium quasiparticles in LCMO, which rapidly interact with phonons before undergoing spin-lattice relaxation on a timescale of tens of picoseconds. This reduces the spin-spin interactions in LCMO, applying stress on BSTO through magnetostriction. This then modifies the FE polarization through the piezoelectric effect, on a timescale much faster than laser-induced heat diffusion from LCMO to BSTO. We have thus demonstrated an ultrafast indirect magnetoelectric effect in a FE/FM heterostructure mediated through elastic coupling, with a timescale primarily governed by spin-lattice relaxation in the FM layer.
\end{abstract}
\maketitle

Magnetoelectric multiferroics have attracted much recent attention, largely due to the scarcity of naturally occurring materials in which magnetic and electric order coexist \cite{Eerenstein2006Nature, Khomskii2009Physics, Cheong2007NM,Ramesh2007NM}. However, the microscopic mechanisms underlying magnetoelectric (ME) coupling between these order parameters have been difficult to unravel, hindering the development of single phase multiferroics with strong ME coupling at useful temperatures \cite{Eerenstein2006Nature,Vaz2010AM,Khomskii2009Physics}. Artificial multiferroic composites \cite{Zheng2004Science,Scott2007Science,Garcia2010Science,Buzzi2013PRL} offer an appealing alternative in which the ME coupling can be engineered through proper choice of the interface geometry and constituent materials \cite{Ramesh2007NM}. The most well known approach for achieving strong ME coupling in these systems uses strain to indirectly couple ferroelectric (FE) and ferromagnetic (FM) order in two-phase layered composites \cite{Scott2007Science,Vaz2010AM,Ma2011AM}. In these composites, the ME effect is essentially a product of the magnetostrictive effect in the FM layer and the piezoelectric effect in the FE layer \cite{Suchtelen1972PRR,Nan2008JAP,Fiebig2009EPJB,Vaz2010AM,Ma2011AM}. A magnetic (B) field is used to modify the spin alignment in the FM layer, which changes the lattice constant through magneto
striction. This in turn applies stress on the FE layer and modifies the FE polarization through the piezoelectric effect \cite{Nan2008JAP,Fiebig2009EPJB,Srinivasan2002PRB}, significantly increasing the ME coupling as compared to single-phase multiferroics \cite{Vaz2010AM,Ma2011AM}. Several reviews describing the ME effect in these composites have been recently published \cite{Nan2008JAP, Vaz2010AM,Ma2011AM, Kambale2011ACMS}.

Despite these impressive advances, an important aspect of multiferroics has received relatively little attention: namely, their dynamic properties  \cite{Khomskii2009Physics,Cheong2007NM,Ma2011AM}, which will impact many of their potential applications. 
In fact, the ultimate timescales limiting the speed of ME coupling in these materials, as well as the physical processses governing them, remain essentially unexplored. Femtosecond optical pulses are particularly useful in this regard \cite{Sheu2012APL,Wen2013PRL,Matsubara2009PRB,Ogawa2009PRB,Fiebig2008JPD,Wang2013ACMS,Hoffmann2011PRB,Jones2014NComm,
Jang2010NJP,Talbayev2008PRL,Kimel2001PRB,Sheu2014PRX,Rana2009AM,Shih2009PRB,Handayani2013JPCM,Qi2012APL}, since they can be used to both interrogate and control the ME response in a non-contact manner on ultrafast timescales, much faster than by directly applying magnetic or electric fields. Furthermore, specific order parameters can be directly accessed using time-resolved second harmonic generation (TR-SHG), making it especially attractive for studying both magnetically \cite{Fiebig2008JPD} and orbitally ordered \cite{Ogawa2013APL} materials as well as multiferroics \cite{Matsubara2009PRB} and interfacial properties of novel layered materials \cite{Ogawa2009PRB} in the time domain.

Here, we use TR-SHG to explore and optically manipulate the coupling between ferroelectric and ferromagnetic order in an oxide heterostructure for the first time, inspired by the strain/stress-based approach described above. By separating the different contributions to the TR-SHG signal in the time domain, we discovered that the timescale dominating the ME response is governed by demagnetization of the FM layer through spin-lattice relaxation. More specifically, optically perturbing magnetic order in the FM layer imposes lateral stress on the FE layer through magnetostriction, modifying FE order within tens of picoseconds (ps). This demonstrates that femtosecond optical pulses can be used not only to give insight into the microscopic mechanisms underlying ME coupling in complex oxide heterostructures, but also to manipulate the ME response in these systems on ultrafast timescales.\\
\\
\textbf{Results}\\
\textbf{Static and time-resolved second harmonic generation (SHG) experiments.} The heterostructure studied here consists of a 50 nm thick film of ferroelectric BSTO and a 50 nm thick film of ferromagnetic LCMO, grown on a MgO substrate; more detail on sample fabrication is given in Methods and in Ref. \cite{Sheu2013PRBR}. We chose BSTO since lattice strain introduced through epitaxial growth can enhance ferroelectricity and the FE transition temperature ($T_{\text{FE}}$), as theoretically predicted \cite{Pertsev1998PRL,Bristowe2012PRB} and experimentally observed \cite{Sheu2013PRBR}. This coupling between strain and FE order makes it a good candidate for inducing indirect ME effects through elastic coupling with a magnetic material. Similarly, LCMO was chosen since a strong ME effect was previously obtained through magnetostriction under an applied DC magnetic field when it was incorporated into a bilayer structure with lead zirconate titanate (PZT) \cite{Srinivasan2002PRB,Vaz2010AM,MEBook}. The concept behind our experiment is thus to demagnetize LCMO with femtosecond optical pulses, modifying its lattice constant through magnetostriction; the resulting stress on BSTO activates its piezoelectric response, changing FE order and indirectly inducing ME coupling within the demagnetization time.
\\
\begin{figure}[tb]
\begin{center}
\includegraphics[width=3.2in]{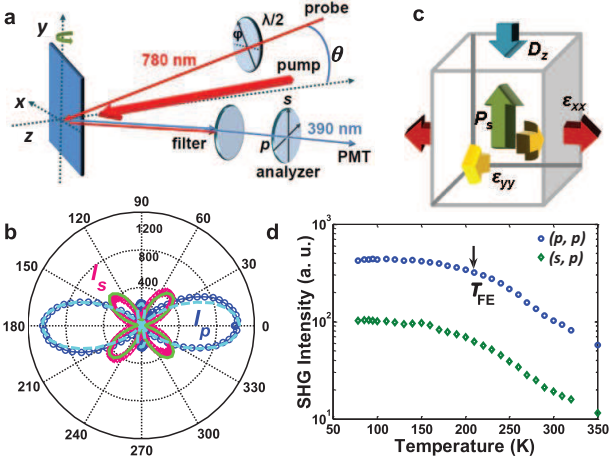}
\caption{ \label{f:concept} \textbf{Experimental setup, static SHG characterization, and piezoelectric response to in-plane stress.} \textbf{(a)} Experimental schematic for TR-SHG in reflection. $\theta$ is the angle between the sample normal and the light propagation direction. $\phi$ is the incident polarization of the fundamental light, which is controlled by a half-wave ($\lambda$/2) plate. Incident $p$ and $s$ polarizations correspond to $\phi$=0$^{\text{o}}$ (180$^{\text{o}}$) and 90$^{\text{o}}$ (270$^{\text{o}}$)   \textbf{(b)} Polar plot of the measured SHG intensity at 10 K. The blue and red symbols are the $p$ and $s$ polarized SHG signals, $I_p$ and $I_s$, respectively, plotted as a function of the incident light polarization. The cyan dashed and green solid lines are numerical fits to $I_p$ and $I_s$ with $d_{31}$=-1, $d_{15}$=-1.2, and $d_{33}$=7.  \textbf{(c)} Schematic diagram of piezoelectric response along $z$ to applied in-plane stress ($\varepsilon_{xx}$ and $\varepsilon_{yy}$).
  \textbf{(d)} Static temperature-dependent SHG for ($p,p$) and ($s,p$) polarization combinations. In LCMO, $T_{\text{C}}\sim$240 K, while in BSTO, $T_{\text{FE}}\sim$215 K \cite{Sheu2013PRBR}, as indicated by the arrow.}
\end{center}
\end{figure}

Photoinduced changes in the static FE polarization of BSTO, $P_s$, can be measured using SHG (Figs. \ref{f:concept}a, \ref{f:concept}b, and \ref{f:concept}d), which is well known to directly probe FE order \cite{Pugachev2012PRL,Miller1964APL,Denev2011SHGRev,Fiebig2005JOSAB}. The SHG polarization generated by the fundamental light fields $E_j$ and $E_k$ is $P_{i}(2\omega)=\epsilon_{0}d_{ijk}E_{j}(\omega)E_{k}(\omega)$, where $\omega$ is the frequency and $d_{ijk}$ is the nonlinear optical susceptibility tensor \cite{Denev2011SHGRev,Fiebig2005JOSAB,Fiebig2002Nature}.  Our BSTO films have $C_{\text{4v}}$ symmetry and a FE polarization along the sample normal ($z$ in Fig. \ref{f:concept}a) due to the in-plane compressive strain induced by lattice matching to LCMO \cite{Sheu2013PRBR}. There are only three non-vanishing tensor components for this crystal symmetry: $d_{15},d_{31}$, and $d_{33}$. By varying the polarization of the incident fundamental and detected SHG light (Fig. \ref{f:concept}a) while measuring the SHG intensity $I$ (Fig. \ref{f:concept}b), these different components can be extracted, as discussed in Methods.

\begin{figure}[tb]
\begin{center}
\includegraphics[width=3.2in]{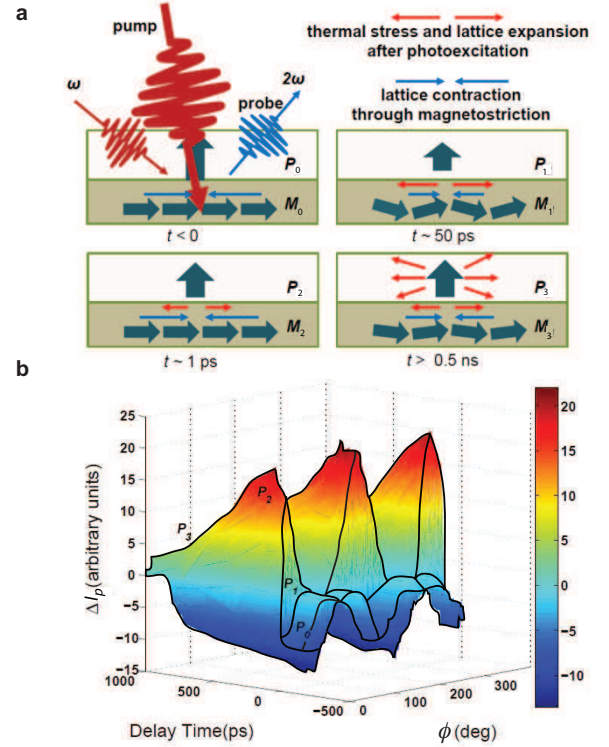}
\caption{ \label{f:3D} \textbf{Photoinduced coupling between ferroelectric and ferromagnetic order.} \textbf{(a)} Schematic diagram of the interplay between the lattice, the magnetization $M$, and the FE polarization $P$ at various pump-probe delays. After photoexcitation, thermal stress on BSTO (thin red arrows) initially results from $e-ph$ coupling in LCMO within 1 ps. Its magnitude gradually increases after relaxation of the lattice contraction in LCMO through magnetostriction (thin blue arrows) on a timescale of $\sim$ 50 ps, causing $P$ to decrease. Thermal diffusion from LCMO to both BSTO and the substrate then takes place at longer timescales, $t>$0.5 ns, finally dissipating through the substrate. \textbf{(b)} Time-resolved $p$-polarized SHG as a function of the incident light polarization at 10 K. $P_{0}$-$P_{3}$ are the FE polarization responses at different times to changes in the magnetization (arising from $M_{0}$-$M_{3}$ in \textbf{(a)}).}
\end{center}
\end{figure}

\begin{figure}[tb]
\begin{center}
\includegraphics[width=3.5in]{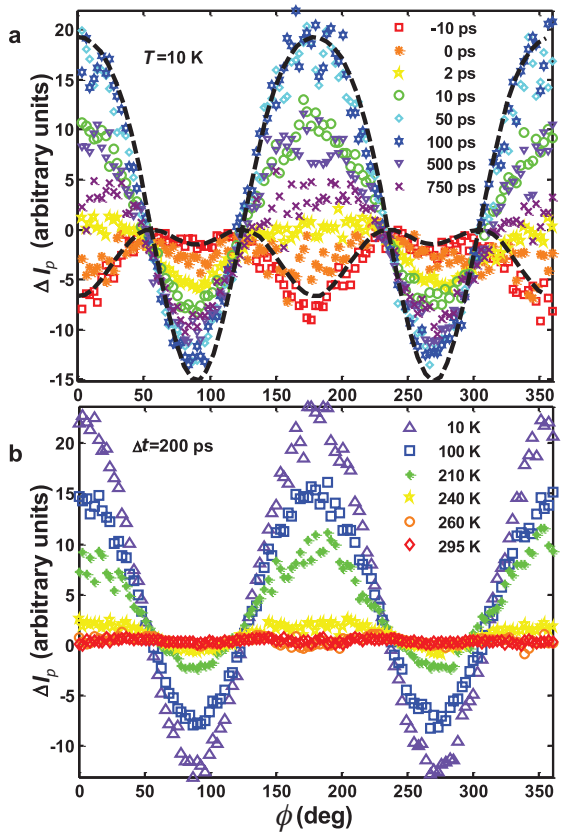}
\caption{\label{f:evidence} \textbf{Evidence for a non-equilibrium, nonlinear origin of the TR-SHG response.}  \textbf{(a)} Polarization-dependent changes in $I_p$ measured at 10 K for various time delays. The data is plotted as symbols and numerical calculations for $t$=-10, 50 ps are plotted as dashed lines. Before $t=0$, the change is due to long-lived residual heating from LCMO to BSTO. After $t=0$ the change arises from reductions in $d_{31}$ and $d_{33}$ (more detail is given in Methods). \textbf{(b)} Polarization-dependent changes in $I_p$ measured at $t$=200 ps for different temperatures.
}
\end{center}
\end{figure}

To optically explore the ME response of our BSTO/LCMO heterostructure, we use the majority of the output laser intensity to photoexcite LCMO (at 1.59 eV, far below the BSTO band gap of $\sim$3.3-3.6 eV) \cite{Sheu2013PRBR}) and the remainder to measure the resulting changes in $P_s$ using SHG. Photoexcited quasiparticles (unbound electrons and holes) in LCMO will relax through different processes that have been extensively studied using pump-probe spectroscopy \cite{Averitt2002JPCM,Muller2009NM,Ogasawara2003PRB}. Initially, the non-equilibrium quasiparticles rapidly lose energy through electron-phonon ($e-ph$) coupling (within $\sim$1 ps), after which the lattice exchanges energy with the spins through spin-lattice ($s-l$) coupling on a timescale of tens to hundreds of ps. On longer timescales ($>$1 nanosecond (ns)) LCMO returns to equilibrium as the remaining energy is dissipated through the substrate. More detail on quasiparticle dynamics in manganites is provided in Supplementary Note 1.\\

\textbf{Data analysis.} We can relate this to the timescales on which the FE polarization changes in our TR-SHG experiments (Fig. \ref{f:3D}). Fig. \ref{f:3D}a schematically shows the different processes occurring for different pump-probe delays $t$. Before $t=0$, a small negative change in the $p$-polarized SHG intensity (\dIp) is observed in all polar combinations at low temperatures ($P_0$), corresponding to residual heating that is not completely dissipated before the arrival of the next pump pulse. This change in $I_{p}$, where the SHG intensity decreases for all incident light polarizations, as more clearly seen in Figs. \ref{f:3D}b and \ref{f:evidence}a for negative time delays, is equivalent to increasing the sample temperature in the SHG polar plot of Fig. \ref{f:concept}b (or the temperature-dependent plots in Fig. \ref{f:concept}d), since before $t$=0 the electron, spin, and lattice subsystems share a common temperature.

\begin{figure}[tb]
\begin{center}
\includegraphics[width=3.5in]{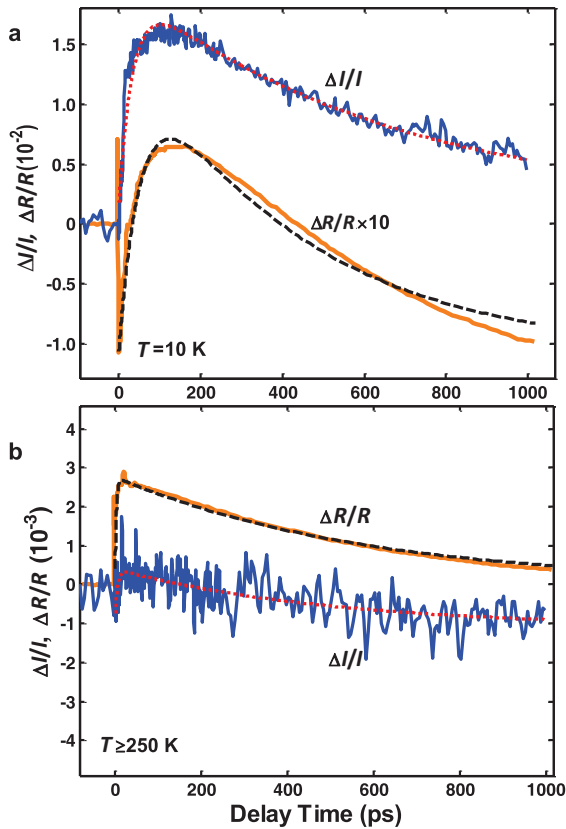}
\caption{ \label{f:dR} \textbf{Comparison of \dR~  and \dI~ at \textbf{(a)} 10 K  and \textbf{(b)} above 250 K.} \dR~ is measured with a $p$-polarized probe and \dI~ is taken in the ($p,p$) configuration. The solid lines are the experimental data and the dashed lines represent a double exponential fit. Additional discussion of the comparison between the \dR~ and \dI~ data is given in Methods and Supplementary Note 1. }
\end{center}
\end{figure}

In contrast, at early times ($t\sim$1-2 ps), we observe a polarization-dependent change in $I_{p}$ ($P_1$), where the SHG intensity both increases and decreases for different polarizations, as shown in Figs. \ref{f:3D}b and Fig. \ref{f:evidence}. This cannot be obtained by simply changing the sample temperature; it is a non-equilibrium effect, occurring before the different subsystems can reach a common temperature. This is due to the elevated temperature of the lattice in LCMO after $e-ph$ coupling, causing it to expand, which in turn initiates an in-plane thermal stress $\varepsilon_{xx}$ and $\varepsilon_{yy}$ on BSTO (without changing its temperature) (Fig. \ref{f:3D}a). This then initiates an out-of-plane piezoelectric response through $D_{i}=e_{ijk}\varepsilon_{jk}$, where $D_{i}$ is the dielectric displacement and $e_{ijk}$ is the piezoelectric coefficient. In the $C_{\text{4v}}$ symmetry, $e_{31}$ is non-zero (more detail is given in Methods), linking $\varepsilon_{xx}$ and $\varepsilon_{yy}$ to $D_z$ (Fig. \ref{f:concept}c) and leading to a photoinduced change in the SHG intensity.  It takes $\sim$7 ps for the strain induced by in-plane stress to propagate through the BSTO layer (estimated from the speed of sound in BSTO), which agrees relatively well with the timescale of the initial rise in \dIp, as shown in Fig. \ref{f:dR} (i.e. there is no rapid $\sim$1 ps rise time, unlike the photoinduced change in reflectivity, \dR). However, the maximum change in $I_p$ (represented by $P_2$) does not occur on this timescale, but on a much slower timescale of $\sim$50-100 ps (Figs. \ref{f:3D}b, and \ref{f:evidence}a).

We can gain insight into the origin of this large change in $I_p$ (and thus $P_s$) by first noting that it primarily occurs below the FM ordering temperature, $T_{\text{C}}\sim$240 K (Fig. \ref{f:evidence}b), indicating that FM order in LCMO is linked to the observed \dIp.  In addition, the timescale is comparable to that for $s-l$ relaxation in LCMO, as determined through comparison of our time-resolved SHG data to optical pump-probe measurements on our BSTO/LCMO heterostructure (Fig. \ref{f:dR}). This suggests that the observed photoinduced ME response is initiated by a change in the lattice constant through magnetostriction \cite{Radaelli1995PRL}, which arises from photoinduced demagnetization of LCMO.  More specifically, $s-l$ relaxation reduces the magnetization in the FM ordered state, gradually decreasing the magnitude of the existing in-plane lattice contraction due to spin-spin interactions  (since the spins in LCMO are aligned in-plane) and applying stress to BSTO (Fig. \ref{f:3D}a, $t\sim$ 50 ps); more detail on general magnetostriction is given in Supplementary Note 2. Then, as for the early time dynamics, the resulting increase in the in-plane tensile stress on BSTO initiates a piezoelectric effect along the $z$ direction (Fig. \ref{f:concept}c), reducing the dielectric displacement $D_z$ and  modifying the FE polarization $P_s$ (Fig. \ref{f:3D}a).

Additional experimental evidence enables us to rule out other possibilities that could lead to the observed change in \dIp. First, \dIp~ at low temperatures does not originate from a photoinduced modulation in the reflectivity due to changes in the optical constants; \dR~ is an order of magnitude smaller than \dI~(the normalized photoinduced change in SHG intensity for a given polarization configuration) at low temperatures (Fig. \ref{f:dR}), and unlike \dI~, it does not vary with the incident probe polarization (Supplementary Figure 1).  Second, there is no observable SHG signal, either static ($I_p$) or photoinduced (\dI), from a single LCMO film, verifying that \dIp~ in our BSTO/LCMO heterostructure does not result from SHG in LCMO. Third, there is no measurable \dI~ signal from a single BSTO film upon 1.59 eV excitation, since the pump is far below the gap \cite{Sheu2013PRBR}.  Fourth, we note that the observed features are independent of the pump polarization, consistent with photoinduced demagnetization in LCMO. Fifth, Kerr rotation of the reflected light at both 1.55 eV and 3.1 eV, induced by demagnetization in LCMO, does not contribute to \dIp. Previous work has shown that the Kerr rotation in manganites for similar pump fluences is $\sim$0.01-0.1 degrees \cite{Ogasawara2003PRB,McGill2005PRB}. This small rotation is unlikely to result in the large changes in SHG intensity that we observed. In addition, if it were to contribute significantly to \dI, it would also be observed in TR-SHG measurements on single LCMO films, where Kerr rotation dominates polarization-dependent measurements \cite{Ogasawara2003PRB,McGill2005PRB}. Moreover, in the unlikely case that Kerr rotation in this heterostructure is larger than that of a single LCMO layer upon photoexcitation, the rotated fundamental or SHG polarizations would induce phase shifts in our polarization-dependent TR-SHG measurements, which are not observed (Fig. \ref{f:3D}b and Fig. \ref{f:evidence}).

Finally, any contributions from surface SHG, including any potential time-dependent interference with the contribution from FE order, are expected to be negligible. The temperature-dependent static SHG signal (Fig. \ref{f:concept}d) originates primarily from bulk FE order in BSTO \cite{Sheu2013PRBR}. Surface and interface contributions from this layer have the same symmetry, $C_{\text{4v}}$, as the bulk FE order. Any interference between these contributions will be either in-phase or 180$^{\text{o}}$ out-of-phase, but will remain constant over the entire sample, as a spatially homogeneous SHG intensity is observed after SHG saturation \cite{Sheu2013PRBR}.  Well above $T_{\text{FE}}$, the residual SHG signal likely originates from the surface or interface and mixes with other potential contributions, such as slow carrier transfer \cite{Sheu2013PRBR} and size effects \cite{Choi2004Science,Pugachev2012PRL} in thin films. Upon photoexcitation, surface SHG does not contribute to the \dI~ signal, since the BSTO surface is not excited. Furthermore, we can exclude any contribution from changes in interfacial SHG due to demagnetization of LCMO, because no TR-SHG signal was obtained from the interface between LCMO and the MgO substrate. These considerations allow us to conclude that the \dI~ signal from our heterostructure originates from changes in $P_s$ that are initiated by photoinduced changes in LCMO.\\
\\
\textbf{Discussion}\\
It is worth re-emphasizing that, as discussed in the introduction, the photoinduced ME response measured here is analogous to the ME effect observed when an external B field is applied to enhance the lattice contraction through magnetostriction in a FE/FM bilayer, inducing a piezoelectric response in the ferroelectric layer \cite{Srinivasan2002PRB}. The major advantage of our approach lies in the temporal degree of freedom offered by femtosecond laser pulses, enabling us to optically manipulate the ME response in the time domain and also discriminate between different contributions to the ME response. For example, we can temporally separate the comparatively small changes in $P_s$ at early times due to heating-induced expansion of LCMO (Fig. \ref{f:3D}, $t\sim$1 ps) from the larger changes in $P_s$ on longer timescales due to the reduction in the LCMO lattice contraction through magnetostriction (Fig. \ref{f:3D}, $t\sim$50 ps). The difference in the magnitude of these changes is likely due to the strong interaction between the ordered spins in LCMO (causing the lattice to contract through magnetostriction), which only permits a small amount of heating-induced thermal stress at early times. As a result, the in-plane tensile stress does not significantly increase before $s-l$ relaxation takes place. Therefore, strain relaxation throughout the entire BSTO film only occurs after tens of ps, setting the timescale for the maximum changes in FE polarization (Fig. \ref{f:evidence}a). Finally, we note that at long (hundreds of ps to ns) timescales, the remaining changes in the FE polarization are due to thermal diffusion from LCMO to BSTO.

In conclusion, we used femtosecond optical pulses to manipulate the coupling between FE and FM order in a complex oxide heterostructure, employing time-resolved SHG to optically perturb the magnetization in LCMO and selectively probe the associated change in the ferroelectric properties of BSTO. We find that although lattice heating in LCMO occurs within $\sim$1 ps, it does not significantly apply lateral stress on BSTO because of the strong lattice contraction originating from spin-spin interactions in LCMO. This lattice contraction is reduced through spin-lattice relaxation, increasing the tensile stress on BSTO through magnetostriction in LCMO and leading to a piezoelectric response that changes the FE polarization. Therefore, coupling between FE and FM order in BSTO/LCMO is induced within tens of picoseconds, mediated through elastic coupling between the BSTO and LCMO layers. Femtosecond optical spectroscopy can thus shed light on the mechanisms underlying ME coupling in complex oxide heterostructures, while presenting intriguing possibilities for future high speed optically controlled magnetoelectric devices (e.g., through coherent spin control \cite{Kimel2005Nature,Kirilyuk2010Review,Bossini2014PRB}, which would enable faster manipulation of spin dynamics and minimize undesired heat flow in device applications.)\\
\\
\textbf{Methods}\\
\textbf{Sample fabrication and characterization.} The samples used in our studies are 50 nm thick BSTO films grown by pulsed laser deposition on 50 nm thick optimally doped manganite films (La$_{0.7}$Ca$_{0.3}$MnO$_{3}$), using (100) MgO substrates. The substrate temperature during film growth is initially optimized and maintained at 750 $^{\text{o}}$C. The oxygen pressure during deposition is 100 mTorr. The samples are cooled to room temperature in pure oxygen (at a pressure of 350 Torr) without further thermal treatment. X-ray diffraction measurements reveal a compressive strain in epitaxially grown BSTO, while LCMO on MgO is relaxed. This causes an increase in tetragonality, even at room temperature, enhancing the FE $T_{\text{C}}$ to $\sim$215 K, which is significantly higher than the bulk $T_{\text{C}}$ $\sim$ 75 K. More detailed characterization of these samples is discussed in Ref. \cite{Sheu2013PRBR}.\\
\\
\textbf{TR-SHG experimental details.} Our time-resolved SHG experiments are based on an amplified Ti:sapphire laser system producing pulses at a 250 kHz repetition rate with $\sim$100 fs duration and energies of $\sim$4 $\mu$J at a center wavelength of 780 nm (1.59 eV). SHG is generated from BSTO at 3.18 eV using a 1.59 eV probe beam, after which it is detected by a photomultiplier tube (PMT) using lock-in detection after filtering out the fundamental signal. The fundamental light polarization is controlled by a half-wave plate ($\lambda$/2 in Fig. \ref{f:concept}a) and the SHG signal is detected for either \textit{p} or \textit{s} polarizations. The standard laser fluence used for SHG in our measurements is $F_0\sim$0.25 mJ cm$^{-2}$ and the pump fluence is $\sim$1 mJ cm$^{-2}$, creating a carrier density of $\sim$10$^{20}$ cm$^{-3}$ ($\sim$10$^{-2}$ per unit cell). In the $C_{\text{4v}}$ symmetry, SHG can be generated and probed when the fundamental light is incident at an angle away from the sample normal, thus exhibiting a sinusoidal dependence on the incident angle $\theta$ (as observed in separate angle-dependent measurements on our samples). Therefore, we used a reflection geometry in which the fundamental light was incident at a $\sim$20$^{\text{o}}$ angle to obtain the data shown in our manuscript. The pump beam was incident at a smaller angle of $\sim$10-15$^{\text{o}}$, and its specular reflection was completely separated from the probe beam path. The temporal delay is achieved by a mechanical delay stage that allows us to vary the probe path length. Finally, ref. \cite{Sheu2013PRBR} gives a detailed static SHG characterization of these samples. Here, all time-resolved data was taken after saturation of the static SHG signal \cite{Sheu2013PRBR}.\\
\\
\textbf{SHG analysis.} For the third rank tensor $d_{ijk}$, the indices $jk$ can be replaced by $l$=1,2,...,6, corresponding to $xx, yy, zz, yz, xz,$ and $xy$, using simple symmetry arguments that reduce
27 tensor components to 18. Further reduction can be achieved through noting that
the $C_{4v}$ symmetry has $d_{31}=d_{zxx}=d_{32}=d_{zyy}$, $d_{15}=d_{xxz}=d_{24}=d_{yyz}$
and  $d_{33}=d_{zzz}$. Therefore, we use the three independent indices 15, 31, and 33 here.
To represent the fourth rank stress tensor $e_{ijk}$, the first $i$=1, 2...,6 as defined above, and
$j,k$=1, 2, 3 correspond to the Cartesian coordinate $x, y, z$. It also can be reduced according to
the above discussion of indexes $j, k$.

Fig. \ref{f:concept}b depicts the static SHG signals for $p$ and $s$ polarizations, $I_p$ and $I_s$. In reflection (Fig. \ref{f:concept}a), $\phi$ determines the polarization of the fundamental light that generates SHG.  With a unit E field of the fundamental light incident at an angle $\theta$ (Fig. \ref{f:concept}a), we have components ($E_{x}(\omega)$,$E_{y}(\omega)$, $E_{z}(\omega)$)=($\cos\phi\cos\theta$,$\sin\phi$,-$\cos\phi\sin\theta$), where $\theta$=0 is along $z$ and $\phi$=0$^{\text{o}}$ and 90$^{\text{o}}$ represent $p$ and $s$ polarizations of the incident fundamental light, respectively. When the detection polarizer is set to $p$ or $s$, we detect the SHG dipole components ($P_x(2\omega)$,$P_y(2\omega)$,$P_z(2\omega)$) along (1,0,1) and (0,1,0), respectively, with each component linked to ($d_{15}=d_{24},d_{31}=d_{32},d_{33}$) by

\begin{eqnarray}
P_x(2\omega)= 2\epsilon_{0}d_{15}E_{x}(\omega)E_{z}(\omega),\\
P_y(2\omega)= 2\epsilon_{0}d_{15}E_{y}(\omega)E_{z}(\omega),\\
P_z(2\omega)= \epsilon_{0}(d_{31}E_{x}^{2}(\omega)+d_{31}E_{y}^{2}(\omega)+d_{33}E_{z}^{2}(\omega)).
\label{e:SHGP}
\end{eqnarray}

We thus can derive the SHG intensities $I_p$ and $I_s$  as functions of $\phi$ (Fig. \ref{f:concept}b):
	
\begin{eqnarray}
\label{e:SHGI}
I_{p}= (a \cos^{2}\phi + b\sin^{2}\phi)^{2},\\
I_{s}= (c \sin(2\phi))^{2},
\end{eqnarray}
where $a,b$ and $c$ are functions of $\theta$ and linear combinations of $d_{15}$, $d_{31}$, and $d_{33}$. This enables us to detect different components of $d_{ijk}$ through different polarization combinations of (incident light, SHG).  The combination of  ($45^{\text{o}}$,$s$) probes $c$, associated with $d_{15}$, while ($s$,$p$) detects $b$, associated with $d_{31}$. The ($p$,$p$) configuration measures $a$ as a linear combination of all three independent components of $d_{ijk}$, allowing us to extract $d_{33}$ (Supplementary Table 1) from the measurements in the ($45^{\text{o}}$,$s$) and ($s$,$p$) configurations.

We can use the above analysis methods to numerically fit our TR-SHG data (Fig. \ref{f:evidence}a). To do this, we need to know the new values of $a$ and $b$ after photoexcitation, which can then be inserted into Eq. (\ref{e:SHGI}). First, we note that the polarization dependence of \dIp~ for $t>0$ (i.e., \dIp~decreases for ($s,p$) and increases for ($p,p$)) (Fig. \ref{f:evidence}) implies that a decrease in $b$ is accompanied by an increase in $a$.  The TR-SHG signal in the ($s$,$p$) configuration is solely due to $b$ (originating from $d_{31}$ (Eq. (\ref{e:SHGI}))),which has the opposite sign of $a$ (Supplementary Table 1).  Therefore, the photoinduced decrease in $d_{31}$ implies an increase in $a$, as observed for the ($p,p$) configuration in Fig. \ref{f:evidence}. Next, we consider the coefficient $c$, which solely originates from $d_{15}$. There are three significant features we notice from the photoinduced change in $d_{15}$. First, \dI~ is much smaller than that measured for the ($p$,$p$) and ($s$,$p$) configurations (Supplementary Figure 1). Second, the change is comparable to the modulation arising from the change in optical constants (Supplementary Figure 1c and Fig. \ref{f:dR}). Third, it does not have significant temperature dependence, unlike the other two polar combinations (Supplementary Figure 1).

We conclude that $d_{15}$ has an insignificant contribution to $a$, and the photoinduced changes in $d_{15}$ could be strongly influenced by photoinduced reflectivity changes in BSTO. These considerations allowed us to simulate the photoinduced changes in $d_{33}$ and $d_{31}$, revealing that $\Delta d_{33}/d_{33}\sim-4\%$ and $\Delta d_{31}/d_{31}\sim-6\%$. We thus can use these reductions in $d_{31}$ and $d_{33}$, together with the isotropic reduction of $d_{31}, d_{33}$ and $d_{15}$ due to residual heating, to numerically fit \dIp~ (dashed lines in Fig. \ref{f:evidence}a). These transient changes lead to the feature observed in $I_p$ after $t_0$=0 (Figs. 3a and Fig. \ref{f:evidence}b).\\
\\

\begin{thebibliography}{50}%
\makeatletter
\providecommand \@ifxundefined [1]{%
 \@ifx{#1\undefined}
}%
\providecommand \@ifnum [1]{%
 \ifnum #1\expandafter \@firstoftwo
 \else \expandafter \@secondoftwo
 \fi
}%
\providecommand \@ifx [1]{%
 \ifx #1\expandafter \@firstoftwo
 \else \expandafter \@secondoftwo
 \fi
}%
\providecommand \natexlab [1]{#1}%
\providecommand \enquote  [1]{``#1''}%
\providecommand \bibnamefont  [1]{#1}%
\providecommand \bibfnamefont [1]{#1}%
\providecommand \citenamefont [1]{#1}%
\providecommand \href@noop [0]{\@secondoftwo}%
\providecommand \href [0]{\begingroup \@sanitize@url \@href}%
\providecommand \@href[1]{\@@startlink{#1}\@@href}%
\providecommand \@@href[1]{\endgroup#1\@@endlink}%
\providecommand \@sanitize@url [0]{\catcode `\\12\catcode `\$12\catcode
  `\&12\catcode `\#12\catcode `\^12\catcode `\_12\catcode `\%12\relax}%
\providecommand \@@startlink[1]{}%
\providecommand \@@endlink[0]{}%
\providecommand \url  [0]{\begingroup\@sanitize@url \@url }%
\providecommand \@url [1]{\endgroup\@href {#1}{\urlprefix }}%
\providecommand \urlprefix  [0]{URL }%
\providecommand \Eprint [0]{\href }%
\providecommand \doibase [0]{http://dx.doi.org/}%
\providecommand \selectlanguage [0]{\@gobble}%
\providecommand \bibinfo  [0]{\@secondoftwo}%
\providecommand \bibfield  [0]{\@secondoftwo}%
\providecommand \translation [1]{[#1]}%
\providecommand \BibitemOpen [0]{}%
\providecommand \bibitemStop [0]{}%
\providecommand \bibitemNoStop [0]{.\EOS\space}%
\providecommand \EOS [0]{\spacefactor3000\relax}%
\providecommand \BibitemShut  [1]{\csname bibitem#1\endcsname}%
\let\auto@bib@innerbib\@empty
\bibitem [{\citenamefont {Eerenstein}\ \emph {et~al.}(2006)\citenamefont
  {Eerenstein}, \citenamefont {Mathur},\ and\ \citenamefont
  {Scott}}]{Eerenstein2006Nature}%
  \BibitemOpen
  \bibfield  {author} {\bibinfo {author} {\bibfnamefont {W.}~\bibnamefont
  {Eerenstein}}, \bibinfo {author} {\bibfnamefont {N.~D.}\ \bibnamefont
  {Mathur}}, \ and\ \bibinfo {author} {\bibfnamefont {J.~F.}\ \bibnamefont
  {Scott}},\ }\bibfield  {title} {\enquote {\bibinfo {title} {Multiferroic and
  magnetoelectric materials},}\ }\href {\doibase 10.1038/nature05023}
  {\bibfield  {journal} {\bibinfo  {journal} {Nature}\ }\textbf {\bibinfo
  {volume} {442}},\ \bibinfo {pages} {759--765} (\bibinfo {year}
  {2006})}\BibitemShut {NoStop}%
\bibitem [{\citenamefont {Khomskii}(2009)}]{Khomskii2009Physics}%
  \BibitemOpen
  \bibfield  {author} {\bibinfo {author} {\bibfnamefont {D.}~\bibnamefont
  {Khomskii}},\ }\bibfield  {title} {\enquote {\bibinfo {title} {Classifying
  multiferroics: Mechanisms and effects},}\ }\href {\doibase
  10.1103/Physics.2.20} {\bibfield  {journal} {\bibinfo  {journal} {Physics}\
  }\textbf {\bibinfo {volume} {2}},\ \bibinfo {pages} {20} (\bibinfo {year}
  {2009})}\BibitemShut {NoStop}%
\bibitem [{\citenamefont {Cheong}\ and\ \citenamefont
  {Mostovoy}(2007)}]{Cheong2007NM}%
  \BibitemOpen
  \bibfield  {author} {\bibinfo {author} {\bibfnamefont {S.-W.}\ \bibnamefont
  {Cheong}}\ and\ \bibinfo {author} {\bibfnamefont {M.}~\bibnamefont
  {Mostovoy}},\ }\bibfield  {title} {\enquote {\bibinfo {title} {Multiferroics:
  a magnetic twist for ferroelectricity},}\ }\href {\doibase 10.1038/nmat1804}
  {\bibfield  {journal} {\bibinfo  {journal} {Nat. Mater.}\ }\textbf {\bibinfo
  {volume} {6}},\ \bibinfo {pages} {13--20} (\bibinfo {year}
  {2007})}\BibitemShut {NoStop}%
\bibitem [{\citenamefont {Ramesh}\ and\ \citenamefont
  {Spaldin}(2007)}]{Ramesh2007NM}%
  \BibitemOpen
  \bibfield  {author} {\bibinfo {author} {\bibfnamefont {R.}~\bibnamefont
  {Ramesh}}\ and\ \bibinfo {author} {\bibfnamefont {N.~A.}\ \bibnamefont
  {Spaldin}},\ }\bibfield  {title} {\enquote {\bibinfo {title} {Multiferroics:
  progress and prospects in thin films},}\ }\href {\doibase 10.1038/nmat1805}
  {\bibfield  {journal} {\bibinfo  {journal} {Nat. Mater.}\ }\textbf {\bibinfo
  {volume} {6}},\ \bibinfo {pages} {21--29} (\bibinfo {year}
  {2007})}\BibitemShut {NoStop}%
\bibitem [{\citenamefont {Vaz}\ \emph {et~al.}(2010)\citenamefont {Vaz},
  \citenamefont {Hoffman}, \citenamefont {Ahn},\ and\ \citenamefont
  {Ramesh}}]{Vaz2010AM}%
  \BibitemOpen
  \bibfield  {author} {\bibinfo {author} {\bibfnamefont {C.~A.~F.}\
  \bibnamefont {Vaz}}, \bibinfo {author} {\bibfnamefont {J.}~\bibnamefont
  {Hoffman}}, \bibinfo {author} {\bibfnamefont {C.~H.}\ \bibnamefont {Ahn}}, \
  and\ \bibinfo {author} {\bibfnamefont {R.}~\bibnamefont {Ramesh}},\
  }\bibfield  {title} {\enquote {\bibinfo {title} {Magnetoelectric coupling
  effects in multiferroic complex oxide composite structures},}\ }\href
  {\doibase 10.1002/adma.200904326} {\bibfield  {journal} {\bibinfo  {journal}
  {Adv. Mater.}\ }\textbf {\bibinfo {volume} {22}},\ \bibinfo {pages}
  {2900--2918} (\bibinfo {year} {2010})}\BibitemShut {NoStop}%
\bibitem [{\citenamefont {Zheng}\ \emph {et~al.}(2004)\citenamefont {Zheng}
  \emph {et~al.}}]{Zheng2004Science}%
  \BibitemOpen
  \bibfield  {author} {\bibinfo {author} {\bibfnamefont {H.}~\bibnamefont
  {Zheng}} \emph {et~al.},\ }\bibfield  {title} {\enquote {\bibinfo {title}
  {Multiferroic {BaTiO$_3$-CoFe$_2$O$_4$} nanostructures},}\ }\href {\doibase
  10.1126/science.1094207} {\bibfield  {journal} {\bibinfo  {journal}
  {Science}\ }\textbf {\bibinfo {volume} {303}},\ \bibinfo {pages} {661--663}
  (\bibinfo {year} {2004})}\BibitemShut {NoStop}%
\bibitem [{\citenamefont {Scott}(2007)}]{Scott2007Science}%
  \BibitemOpen
  \bibfield  {author} {\bibinfo {author} {\bibfnamefont {J.~F.}\ \bibnamefont
  {Scott}},\ }\bibfield  {title} {\enquote {\bibinfo {title} {Applications of
  modern ferroelectrics},}\ }\href {\doibase 10.1126/science.1129564}
  {\bibfield  {journal} {\bibinfo  {journal} {Science}\ }\textbf {\bibinfo
  {volume} {315}},\ \bibinfo {pages} {954--959} (\bibinfo {year}
  {2007})}\BibitemShut {NoStop}%
\bibitem [{\citenamefont {Garcia}\ \emph {et~al.}(2010)\citenamefont {Garcia}
  \emph {et~al.}}]{Garcia2010Science}%
  \BibitemOpen
  \bibfield  {author} {\bibinfo {author} {\bibfnamefont {V.}~\bibnamefont
  {Garcia}} \emph {et~al.},\ }\bibfield  {title} {\enquote {\bibinfo {title}
  {Ferroelectric control of spin polarization},}\ }\href {\doibase
  10.1126/science.1184028} {\bibfield  {journal} {\bibinfo  {journal}
  {Science}\ }\textbf {\bibinfo {volume} {327}},\ \bibinfo {pages} {1106--1110}
  (\bibinfo {year} {2010})}\BibitemShut {NoStop}%
\bibitem [{\citenamefont {Buzzi}\ \emph {et~al.}(2013)\citenamefont {Buzzi},
  \citenamefont {Chopdekar}, \citenamefont {Hockel}, \citenamefont {Bur},
  \citenamefont {Wu}, \citenamefont {Pilet}, \citenamefont {Warnicke},
  \citenamefont {Carman}, \citenamefont {Heyderman},\ and\ \citenamefont
  {Nolting}}]{Buzzi2013PRL}%
  \BibitemOpen
  \bibfield  {author} {\bibinfo {author} {\bibfnamefont {M.}~\bibnamefont
  {Buzzi}}, \bibinfo {author} {\bibfnamefont {R.~V.}\ \bibnamefont
  {Chopdekar}}, \bibinfo {author} {\bibfnamefont {J.~L.}\ \bibnamefont
  {Hockel}}, \bibinfo {author} {\bibfnamefont {A.}~\bibnamefont {Bur}},
  \bibinfo {author} {\bibfnamefont {T.}~\bibnamefont {Wu}}, \bibinfo {author}
  {\bibfnamefont {N.}~\bibnamefont {Pilet}}, \bibinfo {author} {\bibfnamefont
  {P.}~\bibnamefont {Warnicke}}, \bibinfo {author} {\bibfnamefont {G.~P.}\
  \bibnamefont {Carman}}, \bibinfo {author} {\bibfnamefont {L.~J.}\
  \bibnamefont {Heyderman}}, \ and\ \bibinfo {author} {\bibfnamefont
  {F.}~\bibnamefont {Nolting}},\ }\bibfield  {title} {\enquote {\bibinfo
  {title} {Single domain spin manipulation by electric fields in strain coupled
  artificial multiferroic nanostructures},}\ }\href {\doibase
  10.1103/PhysRevLett.111.027204} {\bibfield  {journal} {\bibinfo  {journal}
  {Phys. Rev. Lett.}\ }\textbf {\bibinfo {volume} {111}},\ \bibinfo {pages}
  {027204} (\bibinfo {year} {2013})}\BibitemShut {NoStop}%
\bibitem [{\citenamefont {Ma}\ \emph {et~al.}(2011)\citenamefont {Ma},
  \citenamefont {Hu}, \citenamefont {Li},\ and\ \citenamefont
  {Nan}}]{Ma2011AM}%
  \BibitemOpen
  \bibfield  {author} {\bibinfo {author} {\bibfnamefont {J.}~\bibnamefont
  {Ma}}, \bibinfo {author} {\bibfnamefont {J.}~\bibnamefont {Hu}}, \bibinfo
  {author} {\bibfnamefont {Z.}~\bibnamefont {Li}}, \ and\ \bibinfo {author}
  {\bibfnamefont {C.-W.}\ \bibnamefont {Nan}},\ }\bibfield  {title} {\enquote
  {\bibinfo {title} {Recent progress in multiferroic magnetoelectric
  composites: from bulk to thin films},}\ }\href {\doibase
  10.1002/adma.201003636} {\bibfield  {journal} {\bibinfo  {journal} {Adv.
  Mater.}\ }\textbf {\bibinfo {volume} {23}},\ \bibinfo {pages} {1062--1087}
  (\bibinfo {year} {2011})}\BibitemShut {NoStop}%
\bibitem [{\citenamefont {van Suchtelen}(1972)}]{Suchtelen1972PRR}%
  \BibitemOpen
  \bibfield  {author} {\bibinfo {author} {\bibfnamefont {J.}~\bibnamefont {van
  Suchtelen}},\ }\bibfield  {title} {\enquote {\bibinfo {title} {Product
  properties: A new application of composite materials},}\ }\href@noop {}
  {\bibfield  {journal} {\bibinfo  {journal} {Philips Res. Rep.}\ }\textbf
  {\bibinfo {volume} {27}},\ \bibinfo {pages} {28} (\bibinfo {year}
  {1972})}\BibitemShut {NoStop}%
\bibitem [{\citenamefont {Nan}\ \emph {et~al.}(2008)\citenamefont {Nan},
  \citenamefont {Bichurin}, \citenamefont {Dong}, \citenamefont {Viehland},\
  and\ \citenamefont {Srinivasan}}]{Nan2008JAP}%
  \BibitemOpen
  \bibfield  {author} {\bibinfo {author} {\bibfnamefont {C.-W.}\ \bibnamefont
  {Nan}}, \bibinfo {author} {\bibfnamefont {M.~I.}\ \bibnamefont {Bichurin}},
  \bibinfo {author} {\bibfnamefont {S.}~\bibnamefont {Dong}}, \bibinfo {author}
  {\bibfnamefont {D.}~\bibnamefont {Viehland}}, \ and\ \bibinfo {author}
  {\bibfnamefont {G.}~\bibnamefont {Srinivasan}},\ }\bibfield  {title}
  {\enquote {\bibinfo {title} {Multiferroic magnetoelectric composites:
  Historical perspective, status, and future directions},}\ }\href {\doibase
  http://dx.doi.org/10.1063/1.2836410} {\bibfield  {journal} {\bibinfo
  {journal} {J. Appl. Phys.}\ }\textbf {\bibinfo {volume} {103}},\ \bibinfo
  {pages} {031101} (\bibinfo {year} {2008})}\BibitemShut {NoStop}%
\bibitem [{\citenamefont {Fiebig}\ and\ \citenamefont
  {Spaldin}(2009)}]{Fiebig2009EPJB}%
  \BibitemOpen
  \bibfield  {author} {\bibinfo {author} {\bibfnamefont {M.}~\bibnamefont
  {Fiebig}}\ and\ \bibinfo {author} {\bibfnamefont {N.~A.}\ \bibnamefont
  {Spaldin}},\ }\bibfield  {title} {\enquote {\bibinfo {title} {Current trends
  of the magnetoelectric effect},}\ }\href {\doibase
  10.1140/epjb/e2009-00266-4} {\bibfield  {journal} {\bibinfo  {journal} {Eur.
  Phys. J. B}\ }\textbf {\bibinfo {volume} {71}},\ \bibinfo {pages} {293--297}
  (\bibinfo {year} {2009})}\BibitemShut {NoStop}%
\bibitem [{\citenamefont {Srinivasan}\ \emph {et~al.}(2002)\citenamefont
  {Srinivasan}, \citenamefont {Rasmussen}, \citenamefont {Levin},\ and\
  \citenamefont {Hayes}}]{Srinivasan2002PRB}%
  \BibitemOpen
  \bibfield  {author} {\bibinfo {author} {\bibfnamefont {G.}~\bibnamefont
  {Srinivasan}}, \bibinfo {author} {\bibfnamefont {E.~T.}\ \bibnamefont
  {Rasmussen}}, \bibinfo {author} {\bibfnamefont {B.~J.}\ \bibnamefont
  {Levin}}, \ and\ \bibinfo {author} {\bibfnamefont {R.}~\bibnamefont
  {Hayes}},\ }\bibfield  {title} {\enquote {\bibinfo {title} {Magnetoelectric
  effects in bilayers and multilayers of magnetostrictive and piezoelectric
  perovskite oxides},}\ }\href {\doibase 10.1103/PhysRevB.65.134402} {\bibfield
   {journal} {\bibinfo  {journal} {Phys. Rev. B}\ }\textbf {\bibinfo {volume}
  {65}},\ \bibinfo {pages} {134402} (\bibinfo {year} {2002})}\BibitemShut
  {NoStop}%
\bibitem [{\citenamefont {Kambale}\ \emph {et~al.}(2012)\citenamefont
  {Kambale}, \citenamefont {Jeong},\ and\ \citenamefont
  {Ryu}}]{Kambale2011ACMS}%
  \BibitemOpen
  \bibfield  {author} {\bibinfo {author} {\bibfnamefont {R.~C.}\ \bibnamefont
  {Kambale}}, \bibinfo {author} {\bibfnamefont {D.-Y.}\ \bibnamefont {Jeong}},
  \ and\ \bibinfo {author} {\bibfnamefont {J.}~\bibnamefont {Ryu}},\ }\bibfield
   {title} {\enquote {\bibinfo {title} {Current status of magnetoelectric
  composite thin/thick films},}\ }\href {\doibase 10.1155/2012/824643}
  {\bibfield  {journal} {\bibinfo  {journal} {Adv. Condens. Matter Phys.}\
  }\textbf {\bibinfo {volume} {2012}},\ \bibinfo {pages} {824643} (\bibinfo
  {year} {2012})}\BibitemShut {NoStop}%
\bibitem [{\citenamefont {Sheu}\ \emph {et~al.}(2012)\citenamefont {Sheu},
  \citenamefont {Trugman}, \citenamefont {Park}, \citenamefont {Lee},
  \citenamefont {Yi}, \citenamefont {Cheong}, \citenamefont {Jia},
  \citenamefont {Taylor},\ and\ \citenamefont {Prasankumar}}]{Sheu2012APL}%
  \BibitemOpen
  \bibfield  {author} {\bibinfo {author} {\bibfnamefont {Y.~M.}\ \bibnamefont
  {Sheu}}, \bibinfo {author} {\bibfnamefont {S.~A.}\ \bibnamefont {Trugman}},
  \bibinfo {author} {\bibfnamefont {Y.-S.}\ \bibnamefont {Park}}, \bibinfo
  {author} {\bibfnamefont {S.}~\bibnamefont {Lee}}, \bibinfo {author}
  {\bibfnamefont {H.~T}\ \bibnamefont {Yi}}, \bibinfo {author} {\bibfnamefont
  {S.-W.}\ \bibnamefont {Cheong}}, \bibinfo {author} {\bibfnamefont {Q.~X.}\
  \bibnamefont {Jia}}, \bibinfo {author} {\bibfnamefont {A.~J.}\ \bibnamefont
  {Taylor}}, \ and\ \bibinfo {author} {\bibfnamefont {R.~P.}\ \bibnamefont
  {Prasankumar}},\ }\bibfield  {title} {\enquote {\bibinfo {title} {{Ultrafast
  carrier dynamics and radiative recombination in multiferroic BiFeO$_3$}},}\
  }\href {\doibase 10.1063/1.4729423} {\bibfield  {journal} {\bibinfo
  {journal} {Appl. Phys. Lett.}\ }\textbf {\bibinfo {volume} {100}},\ \bibinfo
  {pages} {242904} (\bibinfo {year} {2012})}\BibitemShut {NoStop}%
\bibitem [{\citenamefont {Wen}\ \emph {et~al.}(2013)\citenamefont {Wen} \emph
  {et~al.}}]{Wen2013PRL}%
  \BibitemOpen
  \bibfield  {author} {\bibinfo {author} {\bibfnamefont {H.}~\bibnamefont
  {Wen}} \emph {et~al.},\ }\bibfield  {title} {\enquote {\bibinfo {title}
  {Electronic origin of ultrafast photoinduced strain in {BiFeO$_{3}$}},}\
  }\href {\doibase 10.1103/PhysRevLett.110.037601} {\bibfield  {journal}
  {\bibinfo  {journal} {Phys. Rev. Lett.}\ }\textbf {\bibinfo {volume} {110}},\
  \bibinfo {pages} {037601} (\bibinfo {year} {2013})}\BibitemShut {NoStop}%
\bibitem [{\citenamefont {Matsubara}\ \emph {et~al.}(2009)\citenamefont
  {Matsubara}, \citenamefont {Kaneko}, \citenamefont {He}, \citenamefont
  {Okamoto},\ and\ \citenamefont {Tokura}}]{Matsubara2009PRB}%
  \BibitemOpen
  \bibfield  {author} {\bibinfo {author} {\bibfnamefont {M.}~\bibnamefont
  {Matsubara}}, \bibinfo {author} {\bibfnamefont {Y.}~\bibnamefont {Kaneko}},
  \bibinfo {author} {\bibfnamefont {J.-P.}\ \bibnamefont {He}}, \bibinfo
  {author} {\bibfnamefont {H.}~\bibnamefont {Okamoto}}, \ and\ \bibinfo
  {author} {\bibfnamefont {Y.}~\bibnamefont {Tokura}},\ }\bibfield  {title}
  {\enquote {\bibinfo {title} {Ultrafast polarization and magnetization
  response of multiferroic {GaFeO$_{3}$} using time-resolved nonlinear optical
  techniques},}\ }\href {\doibase 10.1103/PhysRevB.79.140411} {\bibfield
  {journal} {\bibinfo  {journal} {Phys. Rev. B}\ }\textbf {\bibinfo {volume}
  {79}},\ \bibinfo {pages} {140411} (\bibinfo {year} {2009})}\BibitemShut
  {NoStop}%
\bibitem [{\citenamefont {Ogawa}\ \emph {et~al.}(2009)\citenamefont {Ogawa},
  \citenamefont {Satoh}, \citenamefont {Ogimoto},\ and\ \citenamefont
  {Miyano}}]{Ogawa2009PRB}%
  \BibitemOpen
  \bibfield  {author} {\bibinfo {author} {\bibfnamefont {N.}~\bibnamefont
  {Ogawa}}, \bibinfo {author} {\bibfnamefont {T.}~\bibnamefont {Satoh}},
  \bibinfo {author} {\bibfnamefont {Y.}~\bibnamefont {Ogimoto}}, \ and\
  \bibinfo {author} {\bibfnamefont {K.}~\bibnamefont {Miyano}},\ }\bibfield
  {title} {\enquote {\bibinfo {title} {{Half-metallic spin dynamics at a single
  LaMnO$_{3}$/SrMnO$_{3}$ interface studied with nonlinear magneto-optical Kerr
  effect}},}\ }\href {\doibase 10.1103/PhysRevB.80.241104} {\bibfield
  {journal} {\bibinfo  {journal} {Phys. Rev. B}\ }\textbf {\bibinfo {volume}
  {80}},\ \bibinfo {pages} {241104} (\bibinfo {year} {2009})}\BibitemShut
  {NoStop}%
\bibitem [{\citenamefont {Fiebig}\ \emph {et~al.}(2008)\citenamefont {Fiebig},
  \citenamefont {Duong}, \citenamefont {Satoh}, \citenamefont {Aken},
  \citenamefont {Miyano}, \citenamefont {Tomioka},\ and\ \citenamefont
  {Tokura}}]{Fiebig2008JPD}%
  \BibitemOpen
  \bibfield  {author} {\bibinfo {author} {\bibfnamefont {M.}~\bibnamefont
  {Fiebig}}, \bibinfo {author} {\bibfnamefont {N.~P.}\ \bibnamefont {Duong}},
  \bibinfo {author} {\bibfnamefont {T.}~\bibnamefont {Satoh}}, \bibinfo
  {author} {\bibfnamefont {B.~B.~V.}\ \bibnamefont {Aken}}, \bibinfo {author}
  {\bibfnamefont {K.}~\bibnamefont {Miyano}}, \bibinfo {author} {\bibfnamefont
  {Y.}~\bibnamefont {Tomioka}}, \ and\ \bibinfo {author} {\bibfnamefont
  {Y.}~\bibnamefont {Tokura}},\ }\bibfield  {title} {\enquote {\bibinfo {title}
  {Ultrafast magnetization dynamics of antiferromagnetic compounds},}\
  }\href@noop {} {\bibfield  {journal} {\bibinfo  {journal} {J. Phys. D: Appl.
  Phys.}\ }\textbf {\bibinfo {volume} {41}},\ \bibinfo {pages} {164005}
  (\bibinfo {year} {2008})}\BibitemShut {NoStop}%
\bibitem [{\citenamefont {Wang}\ \emph {et~al.}(2013)\citenamefont {Wang},
  \citenamefont {Luo},\ and\ \citenamefont {Kobayashi}}]{Wang2013ACMS}%
  \BibitemOpen
  \bibfield  {author} {\bibinfo {author} {\bibfnamefont {Y.~T.}\ \bibnamefont
  {Wang}}, \bibinfo {author} {\bibfnamefont {C.~W.}\ \bibnamefont {Luo}}, \
  and\ \bibinfo {author} {\bibfnamefont {T.}~\bibnamefont {Kobayashi}},\
  }\bibfield  {title} {\enquote {\bibinfo {title} {Understanding multiferroic
  hexagonal manganites by static and ultrafast optical spectroscopy},}\ }\href
  {\doibase 10.1155/2013/104806} {\bibfield  {journal} {\bibinfo  {journal}
  {Adv. Condens. Matter Phys.}\ }\textbf {\bibinfo {volume} {2013}},\ \bibinfo
  {pages} {104806} (\bibinfo {year} {2013})}\BibitemShut {NoStop}%
\bibitem [{\citenamefont {Hoffmann}\ \emph {et~al.}(2011)\citenamefont
  {Hoffmann}, \citenamefont {Thielen}, \citenamefont {Becker}, \citenamefont
  {Bohat\'y},\ and\ \citenamefont {Fiebig}}]{Hoffmann2011PRB}%
  \BibitemOpen
  \bibfield  {author} {\bibinfo {author} {\bibfnamefont {T.}~\bibnamefont
  {Hoffmann}}, \bibinfo {author} {\bibfnamefont {P.}~\bibnamefont {Thielen}},
  \bibinfo {author} {\bibfnamefont {P.}~\bibnamefont {Becker}}, \bibinfo
  {author} {\bibfnamefont {L.}~\bibnamefont {Bohat\'y}}, \ and\ \bibinfo
  {author} {\bibfnamefont {M.}~\bibnamefont {Fiebig}},\ }\bibfield  {title}
  {\enquote {\bibinfo {title} {Time-resolved imaging of magnetoelectric
  switching in multiferroic {MnWO$_{4}$}},}\ }\href {\doibase
  10.1103/PhysRevB.84.184404} {\bibfield  {journal} {\bibinfo  {journal} {Phys.
  Rev. B}\ }\textbf {\bibinfo {volume} {84}},\ \bibinfo {pages} {184404}
  (\bibinfo {year} {2011})}\BibitemShut {NoStop}%
\bibitem [{\citenamefont {Jones}\ \emph {et~al.}(2014)\citenamefont {Jones},
  \citenamefont {Gaw}, \citenamefont {Doig}, \citenamefont {Prabhakaran},
  \citenamefont {H\'etroy~Wheeler}, \citenamefont {Boothroyd},\ and\
  \citenamefont {Lloyd-Hughes}}]{Jones2014NComm}%
  \BibitemOpen
  \bibfield  {author} {\bibinfo {author} {\bibfnamefont {S.~P.~P.}\
  \bibnamefont {Jones}}, \bibinfo {author} {\bibfnamefont {S.~M.}\ \bibnamefont
  {Gaw}}, \bibinfo {author} {\bibfnamefont {K.~I.}\ \bibnamefont {Doig}},
  \bibinfo {author} {\bibfnamefont {D.}~\bibnamefont {Prabhakaran}}, \bibinfo
  {author} {\bibfnamefont {E.~M.}\ \bibnamefont {H\'etroy~Wheeler}}, \bibinfo
  {author} {\bibfnamefont {A.~T.}\ \bibnamefont {Boothroyd}}, \ and\ \bibinfo
  {author} {\bibfnamefont {J.}~\bibnamefont {Lloyd-Hughes}},\ }\bibfield
  {title} {\enquote {\bibinfo {title} {High-temperature electromagnons in the
  magnetically induced multiferroic cupric oxide driven by intersublattice
  exchange},}\ }\href {\doibase 10.1038/ncomms4787} {\bibfield  {journal}
  {\bibinfo  {journal} {Nat. Commun.}\ }\textbf {\bibinfo {volume} {5}},\
  \bibinfo {pages} {3787} (\bibinfo {year} {2014})}\BibitemShut {NoStop}%
\bibitem [{\citenamefont {Jang}\ \emph {et~al.}(2010)\citenamefont {Jang},
  \citenamefont {Lim}, \citenamefont {Ahn}, \citenamefont {Kim}, \citenamefont
  {Yee}, \citenamefont {Ahn},\ and\ \citenamefont {Cheong}}]{Jang2010NJP}%
  \BibitemOpen
  \bibfield  {author} {\bibinfo {author} {\bibfnamefont {K.-J.}\ \bibnamefont
  {Jang}}, \bibinfo {author} {\bibfnamefont {J.}~\bibnamefont {Lim}}, \bibinfo
  {author} {\bibfnamefont {J.}~\bibnamefont {Ahn}}, \bibinfo {author}
  {\bibfnamefont {J.-H.}\ \bibnamefont {Kim}}, \bibinfo {author} {\bibfnamefont
  {K.-J.}\ \bibnamefont {Yee}}, \bibinfo {author} {\bibfnamefont {J.~S.}\
  \bibnamefont {Ahn}}, \ and\ \bibinfo {author} {\bibfnamefont {S.-W.}\
  \bibnamefont {Cheong}},\ }\bibfield  {title} {\enquote {\bibinfo {title}
  {{Ultrafast IR spectroscopic study of coherent phonons and dynamic
  spin-lattice coupling in multiferroic LuMnO$_3$}},}\ }\href@noop {}
  {\bibfield  {journal} {\bibinfo  {journal} {New J. Phys.}\ }\textbf {\bibinfo
  {volume} {12}},\ \bibinfo {pages} {023017} (\bibinfo {year}
  {2010})}\BibitemShut {NoStop}%
\bibitem [{\citenamefont {Talbayev}\ \emph {et~al.}(2008)\citenamefont
  {Talbayev}, \citenamefont {Trugman}, \citenamefont {Balatsky}, \citenamefont
  {Kimura}, \citenamefont {Taylor},\ and\ \citenamefont
  {Averitt}}]{Talbayev2008PRL}%
  \BibitemOpen
  \bibfield  {author} {\bibinfo {author} {\bibfnamefont {D.}~\bibnamefont
  {Talbayev}}, \bibinfo {author} {\bibfnamefont {S.~A.}\ \bibnamefont
  {Trugman}}, \bibinfo {author} {\bibfnamefont {A.~V.}\ \bibnamefont
  {Balatsky}}, \bibinfo {author} {\bibfnamefont {T.}~\bibnamefont {Kimura}},
  \bibinfo {author} {\bibfnamefont {A.~J.}\ \bibnamefont {Taylor}}, \ and\
  \bibinfo {author} {\bibfnamefont {R.~D.}\ \bibnamefont {Averitt}},\
  }\bibfield  {title} {\enquote {\bibinfo {title} {Detection of coherent
  magnons via ultrafast pump-probe reflectance spectroscopy in multiferroic
  {${\mathrm{Ba}}_{0.6}{\mathrm{Sr}}_{1.4}{\mathrm{Zn}}_{2}{\mathrm{Fe}}_{12}{\mathrm{O}}_{22}$}},}\
  }\href {\doibase 10.1103/PhysRevLett.101.097603} {\bibfield  {journal}
  {\bibinfo  {journal} {Phys. Rev. Lett.}\ }\textbf {\bibinfo {volume} {101}},\
  \bibinfo {pages} {097603} (\bibinfo {year} {2008})}\BibitemShut {NoStop}%
\bibitem [{\citenamefont {Kimel}\ \emph {et~al.}(2001)\citenamefont {Kimel},
  \citenamefont {Pisarev}, \citenamefont {Bentivegna},\ and\ \citenamefont
  {Rasing}}]{Kimel2001PRB}%
  \BibitemOpen
  \bibfield  {author} {\bibinfo {author} {\bibfnamefont {A.~V.}\ \bibnamefont
  {Kimel}}, \bibinfo {author} {\bibfnamefont {R.~V.}\ \bibnamefont {Pisarev}},
  \bibinfo {author} {\bibfnamefont {F.}~\bibnamefont {Bentivegna}}, \ and\
  \bibinfo {author} {\bibfnamefont {Th.}\ \bibnamefont {Rasing}},\ }\bibfield
  {title} {\enquote {\bibinfo {title} {{Time-resolved nonlinear optical
  spectroscopy of ${\mathrm{Mn}}^{3+}$ ions in rare-earth hexagonal manganites
  $R{\mathrm{MnO}}_{3}$ $(R=\mathrm{Sc},$ Y, Er)}},}\ }\href {\doibase
  10.1103/PhysRevB.64.201103} {\bibfield  {journal} {\bibinfo  {journal} {Phys.
  Rev. B}\ }\textbf {\bibinfo {volume} {64}},\ \bibinfo {pages} {201103}
  (\bibinfo {year} {2001})}\BibitemShut {NoStop}%
\bibitem [{\citenamefont {Sheu}\ \emph {et~al.}(2014)\citenamefont {Sheu},
  \citenamefont {Trugman}, \citenamefont {Yan}, \citenamefont {Qi},
  \citenamefont {Jia}, \citenamefont {Taylor},\ and\ \citenamefont
  {Prasankumar}}]{Sheu2014PRX}%
  \BibitemOpen
  \bibfield  {author} {\bibinfo {author} {\bibfnamefont {Y.~M.}\ \bibnamefont
  {Sheu}}, \bibinfo {author} {\bibfnamefont {S.~A.}\ \bibnamefont {Trugman}},
  \bibinfo {author} {\bibfnamefont {L.}~\bibnamefont {Yan}}, \bibinfo {author}
  {\bibfnamefont {J.}~\bibnamefont {Qi}}, \bibinfo {author} {\bibfnamefont
  {Q.~X.}\ \bibnamefont {Jia}}, \bibinfo {author} {\bibfnamefont {A.~J.}\
  \bibnamefont {Taylor}}, \ and\ \bibinfo {author} {\bibfnamefont {R.~P.}\
  \bibnamefont {Prasankumar}},\ }\bibfield  {title} {\enquote {\bibinfo {title}
  {{Polaronic transport induced by competing interfacial magnetic order in a
  ${\mathrm{La}}_{0.7}{\mathrm{Ca}}_{0.3}{\mathrm{MnO}}_{3}/{\mathrm{BiFeO}}_{3}$
  heterostructure}},}\ }\href {\doibase 10.1103/PhysRevX.4.021001} {\bibfield
  {journal} {\bibinfo  {journal} {Phys. Rev. X}\ }\textbf {\bibinfo {volume}
  {4}},\ \bibinfo {pages} {021001} (\bibinfo {year} {2014})}\BibitemShut
  {NoStop}%
\bibitem [{\citenamefont {Rana}\ \emph {et~al.}(2009)\citenamefont {Rana},
  \citenamefont {Kawayama}, \citenamefont {Mavani}, \citenamefont {Takahashi},
  \citenamefont {Murakami},\ and\ \citenamefont {Tonouchi}}]{Rana2009AM}%
  \BibitemOpen
  \bibfield  {author} {\bibinfo {author} {\bibfnamefont {D.~S.}\ \bibnamefont
  {Rana}}, \bibinfo {author} {\bibfnamefont {I.}~\bibnamefont {Kawayama}},
  \bibinfo {author} {\bibfnamefont {K.}~\bibnamefont {Mavani}}, \bibinfo
  {author} {\bibfnamefont {K.}~\bibnamefont {Takahashi}}, \bibinfo {author}
  {\bibfnamefont {H.}~\bibnamefont {Murakami}}, \ and\ \bibinfo {author}
  {\bibfnamefont {M.}~\bibnamefont {Tonouchi}},\ }\bibfield  {title} {\enquote
  {\bibinfo {title} {Understanding the nature of ultrafast polarization
  dynamics of ferroelectric memory in the multiferroic {BiFeO$_3$}},}\ }\href
  {\doibase 10.1002/adma.200802094} {\bibfield  {journal} {\bibinfo  {journal}
  {Adv. Mater.}\ }\textbf {\bibinfo {volume} {21}},\ \bibinfo {pages}
  {2881--2885} (\bibinfo {year} {2009})}\BibitemShut {NoStop}%
\bibitem [{\citenamefont {Shih}\ \emph {et~al.}(2009)\citenamefont {Shih},
  \citenamefont {Lin}, \citenamefont {Luo}, \citenamefont {Lin}, \citenamefont
  {Uen}, \citenamefont {Juang}, \citenamefont {Wu}, \citenamefont {Lee},
  \citenamefont {Chen},\ and\ \citenamefont {Kobayashi}}]{Shih2009PRB}%
  \BibitemOpen
  \bibfield  {author} {\bibinfo {author} {\bibfnamefont {H.~C.}\ \bibnamefont
  {Shih}}, \bibinfo {author} {\bibfnamefont {T.~H.}\ \bibnamefont {Lin}},
  \bibinfo {author} {\bibfnamefont {C.~W.}\ \bibnamefont {Luo}}, \bibinfo
  {author} {\bibfnamefont {J.-Y.}\ \bibnamefont {Lin}}, \bibinfo {author}
  {\bibfnamefont {T.~M.}\ \bibnamefont {Uen}}, \bibinfo {author} {\bibfnamefont
  {J.~Y.}\ \bibnamefont {Juang}}, \bibinfo {author} {\bibfnamefont {K.~H.}\
  \bibnamefont {Wu}}, \bibinfo {author} {\bibfnamefont {J.~M.}\ \bibnamefont
  {Lee}}, \bibinfo {author} {\bibfnamefont {J.~M.}\ \bibnamefont {Chen}}, \
  and\ \bibinfo {author} {\bibfnamefont {T.}~\bibnamefont {Kobayashi}},\
  }\bibfield  {title} {\enquote {\bibinfo {title} {{Magnetization dynamics and
  the ${\text{Mn}}^{3+}$ $d\text{-}d$ excitation of hexagonal
  ${\text{HoMnO}}_{3}$ single crystals using wavelength-tunable time-resolved
  femtosecond spectroscopy}},}\ }\href {\doibase 10.1103/PhysRevB.80.024427}
  {\bibfield  {journal} {\bibinfo  {journal} {Phys. Rev. B}\ }\textbf {\bibinfo
  {volume} {80}},\ \bibinfo {pages} {024427} (\bibinfo {year}
  {2009})}\BibitemShut {NoStop}%
\bibitem [{\citenamefont {Handayani}\ \emph {et~al.}(2013)\citenamefont
  {Handayani}, \citenamefont {Tobey}, \citenamefont {Janusonis}, \citenamefont
  {Mazurenko}, \citenamefont {Mufti}, \citenamefont {Nugroho}, \citenamefont
  {Tjia}, \citenamefont {Palstra},\ and\ \citenamefont {van
  Loosdrecht}}]{Handayani2013JPCM}%
  \BibitemOpen
  \bibfield  {author} {\bibinfo {author} {\bibfnamefont {I.~P.}\ \bibnamefont
  {Handayani}}, \bibinfo {author} {\bibfnamefont {R.~I.}\ \bibnamefont
  {Tobey}}, \bibinfo {author} {\bibfnamefont {J.}~\bibnamefont {Janusonis}},
  \bibinfo {author} {\bibfnamefont {D.~A.}\ \bibnamefont {Mazurenko}}, \bibinfo
  {author} {\bibfnamefont {N.}~\bibnamefont {Mufti}}, \bibinfo {author}
  {\bibfnamefont {A.~A.}\ \bibnamefont {Nugroho}}, \bibinfo {author}
  {\bibfnamefont {M.~O.}\ \bibnamefont {Tjia}}, \bibinfo {author}
  {\bibfnamefont {T.~T.~M.}\ \bibnamefont {Palstra}}, \ and\ \bibinfo {author}
  {\bibfnamefont {P.~H.~M.}\ \bibnamefont {van Loosdrecht}},\ }\bibfield
  {title} {\enquote {\bibinfo {title} {{Dynamics of photo-excited electrons in
  magnetically ordered TbMnO$_3$}},}\ }\href@noop {} {\bibfield  {journal}
  {\bibinfo  {journal} {J. Phys.: Condens. Matter}\ }\textbf {\bibinfo {volume}
  {25}},\ \bibinfo {pages} {116007} (\bibinfo {year} {2013})}\BibitemShut
  {NoStop}%
\bibitem [{\citenamefont {Qi}\ \emph {et~al.}(2012)\citenamefont {Qi},
  \citenamefont {Yan}, \citenamefont {Zhou}, \citenamefont {Zhu}, \citenamefont
  {Trugman}, \citenamefont {Taylor}, \citenamefont {Jia},\ and\ \citenamefont
  {Prasankumar}}]{Qi2012APL}%
  \BibitemOpen
  \bibfield  {author} {\bibinfo {author} {\bibfnamefont {J.}~\bibnamefont
  {Qi}}, \bibinfo {author} {\bibfnamefont {L.}~\bibnamefont {Yan}}, \bibinfo
  {author} {\bibfnamefont {H.~D.}\ \bibnamefont {Zhou}}, \bibinfo {author}
  {\bibfnamefont {J.-X.}\ \bibnamefont {Zhu}}, \bibinfo {author} {\bibfnamefont
  {S.~A.}\ \bibnamefont {Trugman}}, \bibinfo {author} {\bibfnamefont {A.~J.}\
  \bibnamefont {Taylor}}, \bibinfo {author} {\bibfnamefont {Q.~X.}\
  \bibnamefont {Jia}}, \ and\ \bibinfo {author} {\bibfnamefont {R.~P.}\
  \bibnamefont {Prasankumar}},\ }\bibfield  {title} {\enquote {\bibinfo {title}
  {{Coexistence of coupled magnetic phases in epitaxial TbMnO$_3$ films
  revealed by ultrafast optical spectroscopy}},}\ }\href {\doibase
  http://dx.doi.org/10.1063/1.4754294} {\bibfield  {journal} {\bibinfo
  {journal} {Appl. Phys. Lett.}\ }\textbf {\bibinfo {volume} {101}},\ \bibinfo
  {pages} {122904} (\bibinfo {year} {2012})}\BibitemShut {NoStop}%
\bibitem [{\citenamefont {Ogawa}\ \emph {et~al.}(2013)\citenamefont {Ogawa},
  \citenamefont {Ogimoto},\ and\ \citenamefont {Miyano}}]{Ogawa2013APL}%
  \BibitemOpen
  \bibfield  {author} {\bibinfo {author} {\bibfnamefont {N.}~\bibnamefont
  {Ogawa}}, \bibinfo {author} {\bibfnamefont {Y.}~\bibnamefont {Ogimoto}}, \
  and\ \bibinfo {author} {\bibfnamefont {K.}~\bibnamefont {Miyano}},\
  }\bibfield  {title} {\enquote {\bibinfo {title} {Ultrafast dynamics of
  orbital-order-induced polarization},}\ }\href {\doibase
  http://dx.doi.org/10.1063/1.4812657} {\bibfield  {journal} {\bibinfo
  {journal} {Appl. Phys. Lett.}\ }\textbf {\bibinfo {volume} {102}},\ \bibinfo
  {eid} {251911} (\bibinfo {year} {2013})}\BibitemShut {NoStop}%
\bibitem [{\citenamefont {Sheu}\ \emph {et~al.}(2013)\citenamefont {Sheu},
  \citenamefont {Trugman}, \citenamefont {Yan}, \citenamefont {Chuu},
  \citenamefont {Bi}, \citenamefont {Jia}, \citenamefont {Taylor},\ and\
  \citenamefont {Prasankumar}}]{Sheu2013PRBR}%
  \BibitemOpen
  \bibfield  {author} {\bibinfo {author} {\bibfnamefont {Y.~M.}\ \bibnamefont
  {Sheu}}, \bibinfo {author} {\bibfnamefont {S.~A.}\ \bibnamefont {Trugman}},
  \bibinfo {author} {\bibfnamefont {L.}~\bibnamefont {Yan}}, \bibinfo {author}
  {\bibfnamefont {C.-P.}\ \bibnamefont {Chuu}}, \bibinfo {author}
  {\bibfnamefont {Z.}~\bibnamefont {Bi}}, \bibinfo {author} {\bibfnamefont
  {Q.~X.}\ \bibnamefont {Jia}}, \bibinfo {author} {\bibfnamefont {A.~J.}\
  \bibnamefont {Taylor}}, \ and\ \bibinfo {author} {\bibfnamefont {R.~P.}\
  \bibnamefont {Prasankumar}},\ }\bibfield  {title} {\enquote {\bibinfo {title}
  {{Photoinduced stabilization and enhancement of the ferroelectric
  polarization in
  Ba${}_{0.1}$Sr${}_{0.9}$TiO${}_{3}$/La${}_{0.7}$Ca(Sr)${}_{0.3}$MnO${}_{3}$
  thin film heterostructures}},}\ }\href {\doibase 10.1103/PhysRevB.88.020101}
  {\bibfield  {journal} {\bibinfo  {journal} {Phys. Rev. B}\ }\textbf {\bibinfo
  {volume} {88}},\ \bibinfo {pages} {020101} (\bibinfo {year}
  {2013})}\BibitemShut {NoStop}%
\bibitem [{\citenamefont {Pertsev}\ \emph {et~al.}(1998)\citenamefont
  {Pertsev}, \citenamefont {Zembilgotov},\ and\ \citenamefont
  {Tagantsev}}]{Pertsev1998PRL}%
  \BibitemOpen
  \bibfield  {author} {\bibinfo {author} {\bibfnamefont {N.~A.}\ \bibnamefont
  {Pertsev}}, \bibinfo {author} {\bibfnamefont {A.~G.}\ \bibnamefont
  {Zembilgotov}}, \ and\ \bibinfo {author} {\bibfnamefont {A.~K.}\ \bibnamefont
  {Tagantsev}},\ }\bibfield  {title} {\enquote {\bibinfo {title} {Effect of
  mechanical boundary conditions on phase diagrams of epitaxial ferroelectric
  thin films},}\ }\href {\doibase 10.1103/PhysRevLett.80.1988} {\bibfield
  {journal} {\bibinfo  {journal} {Phys. Rev. Lett.}\ }\textbf {\bibinfo
  {volume} {80}},\ \bibinfo {pages} {1988--1991} (\bibinfo {year}
  {1998})}\BibitemShut {NoStop}%
\bibitem [{\citenamefont {Bristowe}\ \emph {et~al.}(2012)\citenamefont
  {Bristowe}, \citenamefont {Stengel}, \citenamefont {Littlewood},
  \citenamefont {Pruneda},\ and\ \citenamefont {Artacho}}]{Bristowe2012PRB}%
  \BibitemOpen
  \bibfield  {author} {\bibinfo {author} {\bibfnamefont {N.~C.}\ \bibnamefont
  {Bristowe}}, \bibinfo {author} {\bibfnamefont {M.}~\bibnamefont {Stengel}},
  \bibinfo {author} {\bibfnamefont {P.~B.}\ \bibnamefont {Littlewood}},
  \bibinfo {author} {\bibfnamefont {J.~M.}\ \bibnamefont {Pruneda}}, \ and\
  \bibinfo {author} {\bibfnamefont {E.}~\bibnamefont {Artacho}},\ }\bibfield
  {title} {\enquote {\bibinfo {title} {Electrochemical ferroelectric switching:
  Origin of polarization reversal in ultrathin films},}\ }\href {\doibase
  10.1103/PhysRevB.85.024106} {\bibfield  {journal} {\bibinfo  {journal} {Phys.
  Rev. B}\ }\textbf {\bibinfo {volume} {85}},\ \bibinfo {pages} {024106}
  (\bibinfo {year} {2012})}\BibitemShut {NoStop}%
\bibitem [{\citenamefont {Stefanita}(2008)}]{MEBook}%
  \BibitemOpen
  \bibfield  {author} {\bibinfo {author} {\bibfnamefont {C.-G.}\ \bibnamefont
  {Stefanita}},\ }\href@noop {} {\emph {\bibinfo {title} {From Bulk to Nano}}}\
  (\bibinfo  {publisher} {Springer Berlin Heidelberg},\ \bibinfo {year}
  {2008})\BibitemShut {NoStop}%
\bibitem [{\citenamefont {Pugachev}\ \emph {et~al.}(2012)\citenamefont
  {Pugachev}, \citenamefont {Kovalevskii}, \citenamefont {Surovtsev},
  \citenamefont {Kojima}, \citenamefont {Prosandeev}, \citenamefont {Raevski},\
  and\ \citenamefont {Raevskaya}}]{Pugachev2012PRL}%
  \BibitemOpen
  \bibfield  {author} {\bibinfo {author} {\bibfnamefont {A.~M.}\ \bibnamefont
  {Pugachev}}, \bibinfo {author} {\bibfnamefont {V.~I.}\ \bibnamefont
  {Kovalevskii}}, \bibinfo {author} {\bibfnamefont {N.~V.}\ \bibnamefont
  {Surovtsev}}, \bibinfo {author} {\bibfnamefont {S.}~\bibnamefont {Kojima}},
  \bibinfo {author} {\bibfnamefont {S.~A.}\ \bibnamefont {Prosandeev}},
  \bibinfo {author} {\bibfnamefont {I.~P.}\ \bibnamefont {Raevski}}, \ and\
  \bibinfo {author} {\bibfnamefont {S.~I.}\ \bibnamefont {Raevskaya}},\
  }\bibfield  {title} {\enquote {\bibinfo {title} {Broken local symmetry in
  paraelectric {${\mathrm{BaTiO}}_{3}$} proved by second harmonic
  generation},}\ }\href {\doibase 10.1103/PhysRevLett.108.247601} {\bibfield
  {journal} {\bibinfo  {journal} {Phys. Rev. Lett.}\ }\textbf {\bibinfo
  {volume} {108}},\ \bibinfo {pages} {247601} (\bibinfo {year}
  {2012})}\BibitemShut {NoStop}%
\bibitem [{\citenamefont {Miller}(1964)}]{Miller1964APL}%
  \BibitemOpen
  \bibfield  {author} {\bibinfo {author} {\bibfnamefont {R.~C.}\ \bibnamefont
  {Miller}},\ }\bibfield  {title} {\enquote {\bibinfo {title} {Optical second
  harmonic generation in piezoelectric crystals},}\ }\href {\doibase
  10.1063/1.1754022} {\bibfield  {journal} {\bibinfo  {journal} {Appl. Phys.
  Lett.}\ }\textbf {\bibinfo {volume} {5}},\ \bibinfo {pages} {17--19}
  (\bibinfo {year} {1964})}\BibitemShut {NoStop}%
\bibitem [{\citenamefont {Denev}\ \emph {et~al.}(2011)\citenamefont {Denev},
  \citenamefont {Lummen}, \citenamefont {Barnes}, \citenamefont {Kumar},\ and\
  \citenamefont {Gopalan}}]{Denev2011SHGRev}%
  \BibitemOpen
  \bibfield  {author} {\bibinfo {author} {\bibfnamefont {S.~A.}\ \bibnamefont
  {Denev}}, \bibinfo {author} {\bibfnamefont {T.~T.~A.}\ \bibnamefont
  {Lummen}}, \bibinfo {author} {\bibfnamefont {E.}~\bibnamefont {Barnes}},
  \bibinfo {author} {\bibfnamefont {A.}~\bibnamefont {Kumar}}, \ and\ \bibinfo
  {author} {\bibfnamefont {V.}~\bibnamefont {Gopalan}},\ }\bibfield  {title}
  {\enquote {\bibinfo {title} {Probing ferroelectrics using optical second
  harmonic generation},}\ }\href {\doibase 10.1111/j.1551-2916.2011.04740.x}
  {\bibfield  {journal} {\bibinfo  {journal} {J. Am. Ceram. Soc.}\ }\textbf
  {\bibinfo {volume} {94}},\ \bibinfo {pages} {2699} (\bibinfo {year}
  {2011})}\BibitemShut {NoStop}%
\bibitem [{\citenamefont {Fiebig}\ \emph {et~al.}(2005)\citenamefont {Fiebig},
  \citenamefont {Pavlov},\ and\ \citenamefont {Pisarev}}]{Fiebig2005JOSAB}%
  \BibitemOpen
  \bibfield  {author} {\bibinfo {author} {\bibfnamefont {M.}~\bibnamefont
  {Fiebig}}, \bibinfo {author} {\bibfnamefont {V.~V.}\ \bibnamefont {Pavlov}},
  \ and\ \bibinfo {author} {\bibfnamefont {R.~V.}\ \bibnamefont {Pisarev}},\
  }\bibfield  {title} {\enquote {\bibinfo {title} {Second-harmonic generation
  as a tool for studying electronic and magnetic structures of crystals:
  review},}\ }\href {\doibase 10.1364/JOSAB.22.000096} {\bibfield  {journal}
  {\bibinfo  {journal} {J. Opt. Soc. Am. B}\ }\textbf {\bibinfo {volume}
  {22}},\ \bibinfo {pages} {96--118} (\bibinfo {year} {2005})}\BibitemShut
  {NoStop}%
\bibitem [{\citenamefont {Fiebig}\ \emph {et~al.}(2002)\citenamefont {Fiebig},
  \citenamefont {Lottermoser}, \citenamefont {Fr\"{o}hlich}, \citenamefont
  {Goltsev},\ and\ \citenamefont {Pisarev}}]{Fiebig2002Nature}%
  \BibitemOpen
  \bibfield  {author} {\bibinfo {author} {\bibfnamefont {M.}~\bibnamefont
  {Fiebig}}, \bibinfo {author} {\bibfnamefont {Th.}\ \bibnamefont
  {Lottermoser}}, \bibinfo {author} {\bibfnamefont {D.}~\bibnamefont
  {Fr\"{o}hlich}}, \bibinfo {author} {\bibfnamefont {A.~V.}\ \bibnamefont
  {Goltsev}}, \ and\ \bibinfo {author} {\bibfnamefont {R.~V.}\ \bibnamefont
  {Pisarev}},\ }\bibfield  {title} {\enquote {\bibinfo {title} {Observation of
  coupled magnetic and electric domains},}\ }\href {\doibase
  10.1038/nature01077} {\bibfield  {journal} {\bibinfo  {journal} {Nature}\
  }\textbf {\bibinfo {volume} {419}},\ \bibinfo {pages} {818--820} (\bibinfo
  {year} {2002})}\BibitemShut {NoStop}%
\bibitem [{\citenamefont {Averitt}\ and\ \citenamefont
  {Taylor}(2002)}]{Averitt2002JPCM}%
  \BibitemOpen
  \bibfield  {author} {\bibinfo {author} {\bibfnamefont {R.~D.}\ \bibnamefont
  {Averitt}}\ and\ \bibinfo {author} {\bibfnamefont {A.~J.}\ \bibnamefont
  {Taylor}},\ }\bibfield  {title} {\enquote {\bibinfo {title} {Ultrafast
  optical and far-infrared quasiparticle dynamics in correlated electron
  materials},}\ }\href@noop {} {\bibfield  {journal} {\bibinfo  {journal} {J.
  Phys.: Condens. Matter}\ }\textbf {\bibinfo {volume} {14}},\ \bibinfo {pages}
  {R1357} (\bibinfo {year} {2002})}\BibitemShut {NoStop}%
\bibitem [{\citenamefont {M\"{u}ller}\ \emph {et~al.}(2009)\citenamefont
  {M\"{u}ller} \emph {et~al.}}]{Muller2009NM}%
  \BibitemOpen
  \bibfield  {author} {\bibinfo {author} {\bibfnamefont {G.~M.}\ \bibnamefont
  {M\"{u}ller}} \emph {et~al.},\ }\bibfield  {title} {\enquote {\bibinfo
  {title} {Spin polarization in half-metals probed by femtosecond},}\
  }\href@noop {} {\bibfield  {journal} {\bibinfo  {journal} {Nat. Mater.}\
  }\textbf {\bibinfo {volume} {8}},\ \bibinfo {pages} {56--61} (\bibinfo {year}
  {2009})}\BibitemShut {NoStop}%
\bibitem [{\citenamefont {Ogasawara}\ \emph {et~al.}(2003)\citenamefont
  {Ogasawara}, \citenamefont {Matsubara}, \citenamefont {Tomioka},
  \citenamefont {Kuwata-Gonokami}, \citenamefont {Okamoto},\ and\ \citenamefont
  {Tokura}}]{Ogasawara2003PRB}%
  \BibitemOpen
  \bibfield  {author} {\bibinfo {author} {\bibfnamefont {T.}~\bibnamefont
  {Ogasawara}}, \bibinfo {author} {\bibfnamefont {M.}~\bibnamefont
  {Matsubara}}, \bibinfo {author} {\bibfnamefont {Y.}~\bibnamefont {Tomioka}},
  \bibinfo {author} {\bibfnamefont {M.}~\bibnamefont {Kuwata-Gonokami}},
  \bibinfo {author} {\bibfnamefont {H.}~\bibnamefont {Okamoto}}, \ and\
  \bibinfo {author} {\bibfnamefont {Y.}~\bibnamefont {Tokura}},\ }\bibfield
  {title} {\enquote {\bibinfo {title} {{Photoinduced spin dynamics in
  La$_{0.6}$Sr$_{0.4}$MnO$_3$ observed by time-resolved magneto-optical Kerr
  spectroscopy}},}\ }\href {\doibase 10.1103/PhysRevB.68.180407} {\bibfield
  {journal} {\bibinfo  {journal} {Phys. Rev. B}\ }\textbf {\bibinfo {volume}
  {68}},\ \bibinfo {pages} {180407} (\bibinfo {year} {2003})}\BibitemShut
  {NoStop}%
\bibitem [{\citenamefont {Radaelli}\ \emph {et~al.}(1995)\citenamefont
  {Radaelli}, \citenamefont {Cox}, \citenamefont {Marezio}, \citenamefont
  {Cheong}, \citenamefont {Schiffer},\ and\ \citenamefont
  {Ramirez}}]{Radaelli1995PRL}%
  \BibitemOpen
  \bibfield  {author} {\bibinfo {author} {\bibfnamefont {P.~G.}\ \bibnamefont
  {Radaelli}}, \bibinfo {author} {\bibfnamefont {D.~E.}\ \bibnamefont {Cox}},
  \bibinfo {author} {\bibfnamefont {M.}~\bibnamefont {Marezio}}, \bibinfo
  {author} {\bibfnamefont {S~W.}\ \bibnamefont {Cheong}}, \bibinfo {author}
  {\bibfnamefont {P.~E.}\ \bibnamefont {Schiffer}}, \ and\ \bibinfo {author}
  {\bibfnamefont {A.~P.}\ \bibnamefont {Ramirez}},\ }\bibfield  {title}
  {\enquote {\bibinfo {title} {Simultaneous structural, magnetic, and
  electronic transitions in
  {${\mathrm{La}}_{1-x}{\mathrm{Ca}}_{x}\mathrm{Mn}{\mathrm{O}}_{3}$ with
  $x=0.25~\mathrm{and}~0.50$}},}\ }\href {\doibase 10.1103/PhysRevLett.75.4488}
  {\bibfield  {journal} {\bibinfo  {journal} {Phys. Rev. Lett.}\ }\textbf
  {\bibinfo {volume} {75}},\ \bibinfo {pages} {4488--4491} (\bibinfo {year}
  {1995})}\BibitemShut {NoStop}%
\bibitem [{\citenamefont {McGill}\ \emph {et~al.}(2005)\citenamefont {McGill},
  \citenamefont {Miller}, \citenamefont {Torrens}, \citenamefont {Mamchik},
  \citenamefont {Chen},\ and\ \citenamefont {Kikkawa}}]{McGill2005PRB}%
  \BibitemOpen
  \bibfield  {author} {\bibinfo {author} {\bibfnamefont {S.~A.}\ \bibnamefont
  {McGill}}, \bibinfo {author} {\bibfnamefont {R.~I.}\ \bibnamefont {Miller}},
  \bibinfo {author} {\bibfnamefont {O.~N.}\ \bibnamefont {Torrens}}, \bibinfo
  {author} {\bibfnamefont {A.}~\bibnamefont {Mamchik}}, \bibinfo {author}
  {\bibfnamefont {I.-W.}\ \bibnamefont {Chen}}, \ and\ \bibinfo {author}
  {\bibfnamefont {J.~M.}\ \bibnamefont {Kikkawa}},\ }\bibfield  {title}
  {\enquote {\bibinfo {title} {{Optical evidence for transient photoinduced
  magnetization in La$_{0.7}$Ca$_{0.3}$MnO$_3$}},}\ }\href {\doibase
  10.1103/PhysRevB.71.075117} {\bibfield  {journal} {\bibinfo  {journal} {Phys.
  Rev. B}\ }\textbf {\bibinfo {volume} {71}},\ \bibinfo {pages} {075117}
  (\bibinfo {year} {2005})}\BibitemShut {NoStop}%
\bibitem [{\citenamefont {Choi}\ \emph {et~al.}(2004)\citenamefont {Choi},
  \citenamefont {Biegalski}, \citenamefont {Li}, \citenamefont {Sharan},
  \citenamefont {Schubert}, \citenamefont {Uecker}, \citenamefont {Reiche},
  \citenamefont {Chen}, \citenamefont {Pan}, \citenamefont {Gopalan},
  \citenamefont {Chen}, \citenamefont {Schlom},\ and\ \citenamefont
  {Eom}}]{Choi2004Science}%
  \BibitemOpen
  \bibfield  {author} {\bibinfo {author} {\bibfnamefont {K.~J.}\ \bibnamefont
  {Choi}}, \bibinfo {author} {\bibfnamefont {M.}~\bibnamefont {Biegalski}},
  \bibinfo {author} {\bibfnamefont {Y.~L.}\ \bibnamefont {Li}}, \bibinfo
  {author} {\bibfnamefont {A.}~\bibnamefont {Sharan}}, \bibinfo {author}
  {\bibfnamefont {J.}~\bibnamefont {Schubert}}, \bibinfo {author}
  {\bibfnamefont {R.}~\bibnamefont {Uecker}}, \bibinfo {author} {\bibfnamefont
  {P.}~\bibnamefont {Reiche}}, \bibinfo {author} {\bibfnamefont {Y.~B.}\
  \bibnamefont {Chen}}, \bibinfo {author} {\bibfnamefont {X.~Q.}\ \bibnamefont
  {Pan}}, \bibinfo {author} {\bibfnamefont {V.}~\bibnamefont {Gopalan}},
  \bibinfo {author} {\bibfnamefont {L.-Q.}\ \bibnamefont {Chen}}, \bibinfo
  {author} {\bibfnamefont {D.~G.}\ \bibnamefont {Schlom}}, \ and\ \bibinfo
  {author} {\bibfnamefont {C.~B.}\ \bibnamefont {Eom}},\ }\bibfield  {title}
  {\enquote {\bibinfo {title} {{Enhancement of ferroelectricity in strained
  BaTiO$_3$ thin films}},}\ }\href {\doibase 10.1126/science.1103218}
  {\bibfield  {journal} {\bibinfo  {journal} {Science}\ }\textbf {\bibinfo
  {volume} {306}},\ \bibinfo {pages} {1005--1009} (\bibinfo {year}
  {2004})}\BibitemShut {NoStop}%
\bibitem [{\citenamefont {Kimel}\ \emph {et~al.}(2005)\citenamefont {Kimel},
  \citenamefont {Kirilyuk}, \citenamefont {Usachev}, \citenamefont {Pisarev},
  \citenamefont {Balbashov},\ and\ \citenamefont {Rasing}}]{Kimel2005Nature}%
  \BibitemOpen
  \bibfield  {author} {\bibinfo {author} {\bibfnamefont {A.~V.}\ \bibnamefont
  {Kimel}}, \bibinfo {author} {\bibfnamefont {A.}~\bibnamefont {Kirilyuk}},
  \bibinfo {author} {\bibfnamefont {P.~A.}\ \bibnamefont {Usachev}}, \bibinfo
  {author} {\bibfnamefont {R.~V.}\ \bibnamefont {Pisarev}}, \bibinfo {author}
  {\bibfnamefont {A.~M.}\ \bibnamefont {Balbashov}}, \ and\ \bibinfo {author}
  {\bibfnamefont {Th.}\ \bibnamefont {Rasing}},\ }\bibfield  {title} {\enquote
  {\bibinfo {title} {Ultrafast non-thermal control of magnetization by
  instantaneous photomagnetic pulses},}\ }\href {\doibase 10.1038/nature03564}
  {\bibfield  {journal} {\bibinfo  {journal} {Nature}\ }\textbf {\bibinfo
  {volume} {435}},\ \bibinfo {pages} {655--657} (\bibinfo {year}
  {2005})}\BibitemShut {NoStop}%
\bibitem [{\citenamefont {Kirilyuk}\ \emph {et~al.}(2010)\citenamefont
  {Kirilyuk}, \citenamefont {Kimel},\ and\ \citenamefont
  {Rasing}}]{Kirilyuk2010Review}%
  \BibitemOpen
  \bibfield  {author} {\bibinfo {author} {\bibfnamefont {A.}~\bibnamefont
  {Kirilyuk}}, \bibinfo {author} {\bibfnamefont {A.~V.}\ \bibnamefont {Kimel}},
  \ and\ \bibinfo {author} {\bibfnamefont {Th.}\ \bibnamefont {Rasing}},\
  }\bibfield  {title} {\enquote {\bibinfo {title} {Ultrafast optical
  manipulation of magnetic order},}\ }\href {\doibase
  10.1103/RevModPhys.82.2731} {\bibfield  {journal} {\bibinfo  {journal} {Rev.
  Mod. Phys.}\ }\textbf {\bibinfo {volume} {82}},\ \bibinfo {pages}
  {2731--2784} (\bibinfo {year} {2010})}\BibitemShut {NoStop}%
\bibitem [{\citenamefont {Bossini}\ \emph {et~al.}(2014)\citenamefont
  {Bossini}, \citenamefont {Kalashnikova}, \citenamefont {Pisarev},
  \citenamefont {Rasing},\ and\ \citenamefont {Kimel}}]{Bossini2014PRB}%
  \BibitemOpen
  \bibfield  {author} {\bibinfo {author} {\bibfnamefont {D.}~\bibnamefont
  {Bossini}}, \bibinfo {author} {\bibfnamefont {A.~M.}\ \bibnamefont
  {Kalashnikova}}, \bibinfo {author} {\bibfnamefont {R.~V.}\ \bibnamefont
  {Pisarev}}, \bibinfo {author} {\bibfnamefont {Th.}\ \bibnamefont {Rasing}}, \
  and\ \bibinfo {author} {\bibfnamefont {A.~V.}\ \bibnamefont {Kimel}},\
  }\bibfield  {title} {\enquote {\bibinfo {title} {Controlling coherent and
  incoherent spin dynamics by steering the photoinduced energy flow},}\ }\href
  {\doibase 10.1103/PhysRevB.89.060405} {\bibfield  {journal} {\bibinfo
  {journal} {Phys. Rev. B}\ }\textbf {\bibinfo {volume} {89}},\ \bibinfo
  {pages} {060405} (\bibinfo {year} {2014})}\BibitemShut {NoStop}%
\end{thebibliography}
%
\textbf{Acknowledgements} This work was supported by the U. S. Department of Energy, Office of Basic Energy Sciences, Division of Material Sciences and Engineering. The samples were fabricated at the Center for Integrated Nanotechnologies, a U. S. Department of Energy, Office of Basic Energy Sciences user facility. Partial support was also provided by Los Alamos National Laboratory's Directed Research and Development program. Los Alamos National Laboratory, an affirmative action equal opportunity employer, is operated by Los Alamos National Security, LLC, for the National Nuclear Security Administration of the US Department of Energy under Contract No. DE-AC52-06NA25396.\\
\\
\textbf{Author contributions} Y.M.S. designed and carried out the experiments and data analysis, with input from R.P.P. and A.J.T. Y.M.S. and S.A.T. performed the theoretical analysis and physical interpretation of the data, with additional input from R.P.P. Y.M.S. and R.P.P. wrote the manuscript, with additional input from all coauthors. R.P.P. and A.J.T. directed the project. Y.L. and Q.X.J. prepared the samples for the experiments.\\
\\
\\

\end{document}